\documentclass[journal]{IEEEtran}
\usepackage{cite}
\usepackage{pifont}

\ifCLASSINFOpdf
   \usepackage[pdftex]{graphicx}
   \DeclareGraphicsExtensions{.pdf,.jpeg,.png}
\else
   \usepackage[dvips]{graphicx}
\fi

\usepackage{lipsum}

\usepackage{amsmath}
\usepackage{algorithmic}
\usepackage{array}

\ifCLASSOPTIONcompsoc
 \usepackage[caption=false,font=normalsize,labelfont=sf,textfont=sf]{subfig}
\else
 \usepackage[caption=false,font=footnotesize]{subfig}
\fi
\usepackage{dblfloatfix}

\usepackage{xcolor}

\newcommand\revised[1]{\textcolor{black}{#1}}
\newcommand\revisedd[1]{\textcolor{black}{#1}}
\newcommand\reviseddd[1]{\textcolor{black}{#1}}

\ifCLASSOPTIONcaptionsoff
 \usepackage[nomarkers]{endfloat}
\let\MYoriglatexcaption\caption
\renewcommand{\caption}[2][\relax]{\MYoriglatexcaption[#2]{#2}}
\fi

\usepackage{url}
\usepackage{amsfonts} 
\usepackage{adjustbox}
\usepackage{graphicx}

\usepackage{todonotes}
\usepackage{subfiles}
\definecolor{startpoint}{rgb}{60,120,216}

\begin{document}
\title{Low-Cost Superconducting Fan-Out \\with \revisedd{Cell \reviseddd{I$_\text{C}$ Ranking}}}

\author{Jennifer~Volk,
        Georgios~Tzimpragos, Alex~Wynn, Evan~Golden
        and~Timothy~Sherwood
\thanks{J. Volk is in the Department
of Electrical and Computer Engineering at the University of California, Santa Barbara, CA, 93106 USA. e-mail: jevolk@ucsb.edu}
\thanks{G. Tzimpragos is in the Electrical Engineering and Computer Science Department at the University of Michigan, Ann Arbor, MI, 48109 USA.}
\thanks{A. Wynn and E. Golden are at the Massachusetts Institute of Technology Lincoln Laboratory, Lexington, MA 02421 USA.}%
\thanks{T. Sherwood is in the Computer Science Department at the University of California, Santa Barbara.}
}

\maketitle

\begin{abstract}

Superconductor electronics (SCE) promise computer systems with orders of magnitude higher speeds and lower energy consumption than their complementary metal-oxide semiconductor (CMOS) counterparts. At the same time, the scalability and \revised{resource} utilization of superconducting systems are major concerns. Some of these concerns come from device-level challenges and the gap between SCE and CMOS technology nodes, and others come from the way Josephson Junctions (JJs) are used. Towards this end, we notice that a considerable fraction of hardware resources are not involved in logic operations, but rather are used for fan-out and buffering purposes. In this paper, we ask if there is a way to reduce these overheads, propose the \reviseddd{use} of JJs at the cell boundaries \reviseddd{to increase the number of outputs that a single stage can drive}, and establish a set of rules to discretize critical currents in a way that is conducive to this assignment. Finally, we \revisedd{explore the design trade-offs that the presented approach opens up and demonstrate its promise} through detailed analog simulations and modeling analyses. \revised{Our experiments indicate that the introduced method leads to a 48\% savings in the JJ count for a tree with a fan-out of 1024, as well as an average of 43\% of the JJ count for signal splitting and 32\% for clock \reviseddd{splitting} in ISCAS'85 benchmarks.}
\end{abstract}

\section{Introduction}

\IEEEPARstart{T}{he} performance and energy characteristics of superconductor electronics (SCE) put them in the spotlight as prime candidates for large-scale computing \cite{holmes2017superconductor}, quantum computing~\cite{li2019fabrication}, and machine learning~\cite{feldhoff2021niobium, he2021design,supernpu}. Their low computational density, however, is still a major roadblock ahead \cite{tolpygo2016superconductor,tolpygo2020increasing}. That said, the computational density of SCE systems is predominantly a function of: the technology node, which governs the area of Josephson Junctions (JJs) and transmission lines; the total number of JJs needed to implement the desired function; and the total number of required transmission lines \cite{tolpygo2018superconductor}. In that regard, device researchers focus on the miniaturization of existing technology nodes \cite{mukhanov1997josephson,tolpygo2017properties} while design automation experts concentrate on logic optimizations~\cite{fourie2020electronic} and architects on joint logic and microarchitectural optimizations~\cite{Tzimpragos2021SuperconductingCW,tzimpragos2021temporal,10.1145/3373376.3378517}. In this paper, we adopt a different perspective on this problem by instead looking at reducing the number of JJs used for electrical purposes.

A close look at existing Single Flux Quantum (SFQ)\footnote{SFQ is used here to describe the Rapid Single Flux Quantum (RSFQ) logic and its variants such as ERSFQ, eSFQ, etc.~\cite{ultimateperf2,ersfq1,esfq}.} benchmark designs reveals that 15-33\% of the total JJ count is taken up by splitters~\cite{pedram_superconductive_2020,cong_8-b_2021}. Splitters act as amplifying Josephson Transmission Lines (JTLs) that deliver two copies of an input SFQ to its outputs and are essential for fan-out in SFQ circuit design~\cite{likharev_semenov91}. On top of their hardware cost, multiple JJs per logic cell are typically used for buffering purposes, accounting for an additional $\sim$20\% of the JJ count~\cite{coldfluxlib,polonskyOR}. 

The goal of this paper is to reduce this overhead and make room for more logic cells and greater functionality within the same area. We observe that the JJs used for splitting and buffering perform related roles, and can be ``merged'' by utilizing just one set of JJs to perform both tasks. In other words, we can integrate the capacity for fan-out\reviseddd{---a term that is used in this work to refer to the number of outputs of a cell, irrespective of JJ sizes---}into the logic cells themselves through a novel JJ \revisedd{sharing} technique. \reviseddd{JJ sharing alludes to the process of replacing JJs in a splitter cell with JJs from surrounding cells}.  
\revised{To achieve \reviseddd{this integration}, JJ sizes must be tuned on a case-by-case basis to match the desired critical currents (I$_\text{C}$s) and meet fan-out needs---a process that is time-consuming at medium to large scales and can be prone to errors, especially under the complex interconnection properties of SFQ cells~\cite{tolpygo2016superconductor,deng1997data}.} We propose a new abstraction\reviseddd{, dubbed cell I$_\text{C}$ ranking\footnote{\reviseddd{For conciseness, we refer to cell I$_\text{C}$ ranking as cell ranking from here on.}},} to simplify the design of multiple-fan-out SFQ circuits \reviseddd{through a fan-out-aware assignment of discrete baseline I$_\text{C}$s to cells.} \revisedd{In this context, a cell's \reviseddd{baseline I$_\text{C}$} is defined as the I$_\text{C}$ of its buffer JJs \reviseddd{and serves as a parametric scaling factor for all other JJs, inductors, and resistors in the cell~\cite{9681154}}}. We also derive a set of guidelines to systematically generate these assignments \revised{and provide a good starting point for further optimizations}. 

\revised{To validate this approach, we \revisedd{analyze how metrics such as area, energy, and delay are affected by the way cell ranks are assigned and} demonstrate functional correctness, wide bias margins, and improved resource efficiency\footnote{\revisedd{For accurate area results, complete circuit layouts are necessary. However, considering that a greater number of JJs in a design also implies a greater number of inductors and resistors, JJ count is commonly used as a first-order estimate of resource efficiency\cite{irds22}.}} for various example designs through a combination of detailed SPICE simulations\footnote{\revised{Our design netlists are available at \url{https://github.com/UCSBarchlab/SFQ-Ranking.}}}
and higher-level models.} Particularly, JJ \revisedd{sharing} enables a fan-out of 8 in a single stage with the use of JJs at the edges of cells---boundary JJs---for splitting with \revised{$\pm37\%$} bias margins on average. The application of the introduced design rules leads to about 20\% savings in the total JJ count for a simulated 2-bit Kogge-Stone Adder (KSA) design. \revised{JJ sharing and cell ranking} \revised{also give an estimated average savings of 10\% in the total JJ count within ISCAS’85 benchmarks\revised{~\cite{brglez1985neutral}}, \cite{hansen1999unveiling}---43\% of the JJ count for signal splitting and 32\% for clock fan-out---even without taking into consideration the effects of path-balancing, and 47\% for a tree with fan-out of 1024 (FO1024) compared to the use of conventional splitters.}

Overall, the main contributions of this paper are: (1) the \revisedd{use} of boundary JJs for splitting purposes; (2) the abstraction of critical currents through cell ranking; (3) a set of rules that allows modular design at the gate level while enforcing circuit-level constraints; \revisedd{and (4) an analysis of cell rank as a new design trade-off parameter}.

\begin{figure}
    \centering
    \includegraphics[scale=0.45]{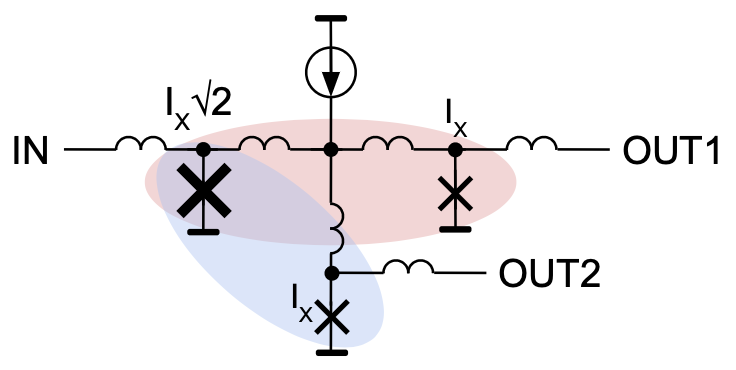}
    \caption{A Single Flux Quantum (SFQ) active splitter consists of three Josephson Junctions (JJs) and drives a fan-out of 2 (FO2). It does not perform any logic operation but rather acts as two Josephson Transmission Lines (JTLs), here red and blue, by directly passing SFQ from its input to both of its outputs. The sum of the critical currents of the two output JJs is 2I$_\text{X}$, $\sqrt{2}$ times larger than that of the input, so the whole element is amplifying. Splitters can be chained directly, one after the other, with no degradation in bias margins.}
    \label{fig:splitter}
\end{figure}

\section{Reducing Electrical Redundancies}
\subsection{The Splitting Problem}
The poor scalability of active splitters, which are traditionally used to replicate SFQ pulses~\cite{likharev_semenov91}, has long been recognized as one of the key challenges in SCE~\cite{smart21,fourieoverview18,friedman17,ptlwiring07}. Moreover, high fan-out needs~\cite{schneider_fan-out_2020}, especially in the case of conventional fully-synchronous SFQ designs~\cite{friedman_htree, wang2021novel, jabbari2020global}, compound this problem. For example, in the FLUX processor, 77\% of the instruction memory area goes to splitters, mergers, and transmission lines for address decoding~\cite{fluxcoredesign01}. A synthesized 8-bit general purpose RISC processor uses 21,065 splitters for clock distribution, accounting for nearly 16\% of the total 400,000 JJ count---and this does not include splitting on the data path~\cite{kameda2007new}. Moreover, a synthesized 4-bit KSA uses 58 clock splitters and 31 signal splitters, a cell count that exceeds the number of logic cells and path delay D-flip flops by 54\%~\cite{fourie2018extraction}. Obviously, if we could minimize these splitting overheads, we would open a significant amount of space for additional logic.

Before discussing splitting optimizations, let's take a look at the construction of a splitter to understand its functionality and constraints. \revised{Fig.}~\ref{fig:splitter} depicts the corresponding schematic. As can be seen, three JJs are needed to achieve an FO2. Each of the two legs of the splitter, colored in red and blue, acts as a JTL. The critical current of the input JJ is $\sqrt{2}$ times larger than the critical currents of the output JJs~\cite{likharev_semenov91}, the latter of which match the baseline critical currents of the preceding and succeeding SFQ cells. The difference in critical currents makes the entire splitter act as an amplifying JTL. In other words, the larger input JJ boosts incoming SFQ pulses so that they still meet the current requirements after the fan-out juncture to switch the smaller-sized output JJs. 

Supporting a fan-out of $N$ typically implies the use of a splitter tree that consists of $(N-1)$ FO2 splitters and at least $log_2(N)$ stages. One way to improve this situation is by increasing the number of output ports per splitter cell~\cite{massoudmultifanout}. Another improvement comes from a recently-proposed JJ sharing policy in which splitter output JJs are reused as input JJs for the next splitters~\cite{jabbari_splitter_2021}. In the case where splitters are followed by passive transmission lines (PTLs), further delay and area gains can be achieved by moving these PTLs from the outputs of the splitter cell to its center to replace the shared JJs~\cite{yamada2006novel}. These approaches are promising but require careful tuning of JJ critical currents, as well as the use of amplifying JTLs on their inputs and outputs for smooth integration with the rest of the circuit. However, replacing splitters with amplifying JTLs~\revisedd{as a one-for-one substitution} can hamper the gains made by the elimination of splitters. The goal of the proposed work is to instead extend the functionality of JJs within logic cells. We achieve this \revisedd{by applying JJ sharing to SFQ logic cells for the first time and} introducing a systematic way to manage the \revisedd{shared} JJs. \reviseddd{In doing so, we reduce the cost of fan-out}.

\subsection{JJ Characterization}

\begin{figure}
    \centering
    \includegraphics[scale=0.125]{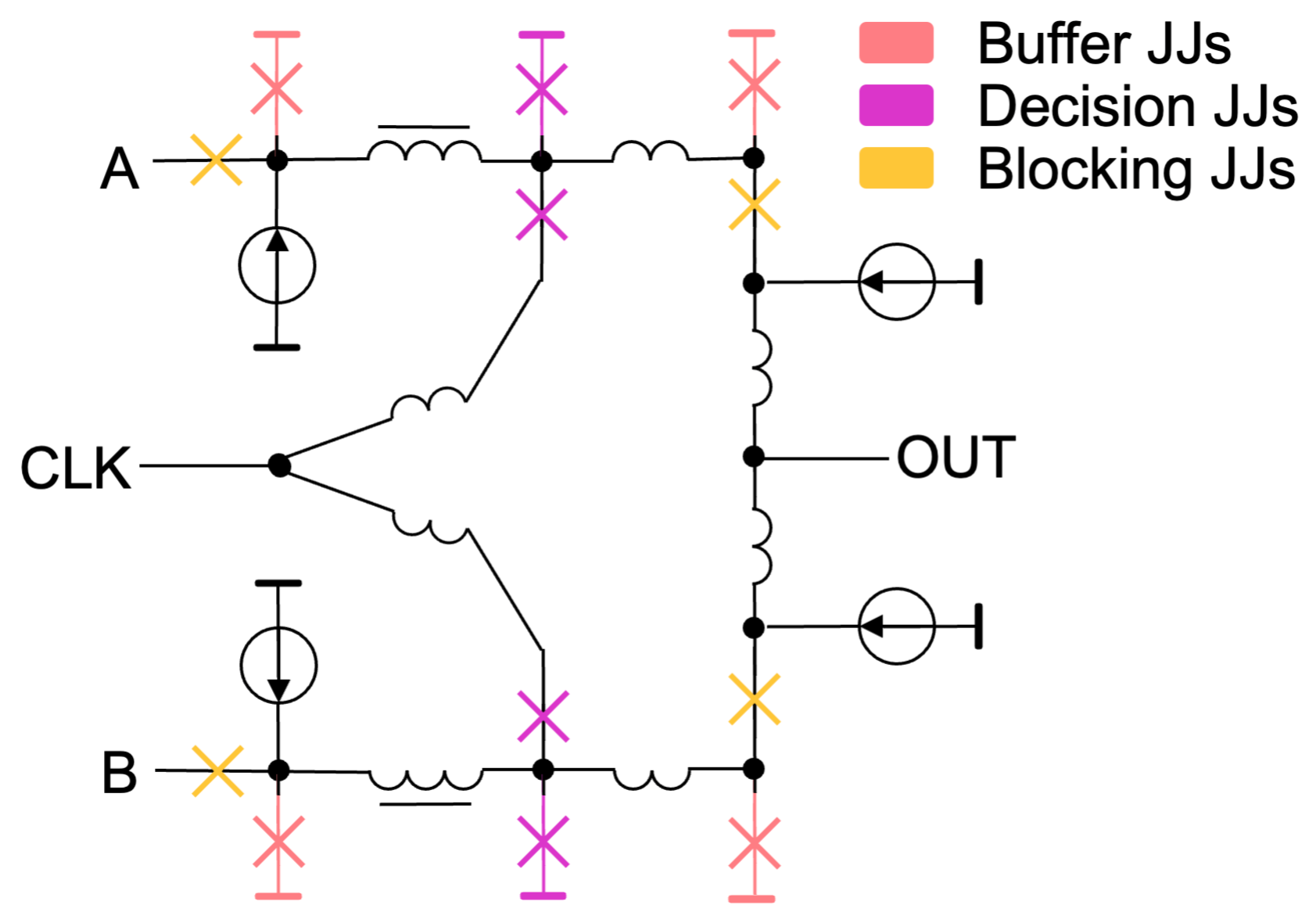}
    \caption{An example SFQ OR cell, in which each JJ is color-coded to reflect its function in the cell. \revised{Pink} JJs buffer input data, and in doing so help maintain signal fidelity. \revised{Magenta} JJs, indicating decision-making pairs, serve as comparators that evaluate the desired logical function. Lastly, \revised{gold} blocking JJs are connected serially and perform two tasks: when they precede a Superconducting Quantum Interference Device (SQUID) in which an SFQ is currently stored, they prevent additional SFQ from entering the loop; and when they follow a SQUID or are close to the cell output, they prevent the backward propagation of SFQ from the output line towards the opposite input.}
    \label{fig:orgate}
\end{figure}

\begin{figure}
    \centering
    \includegraphics[scale=0.17]{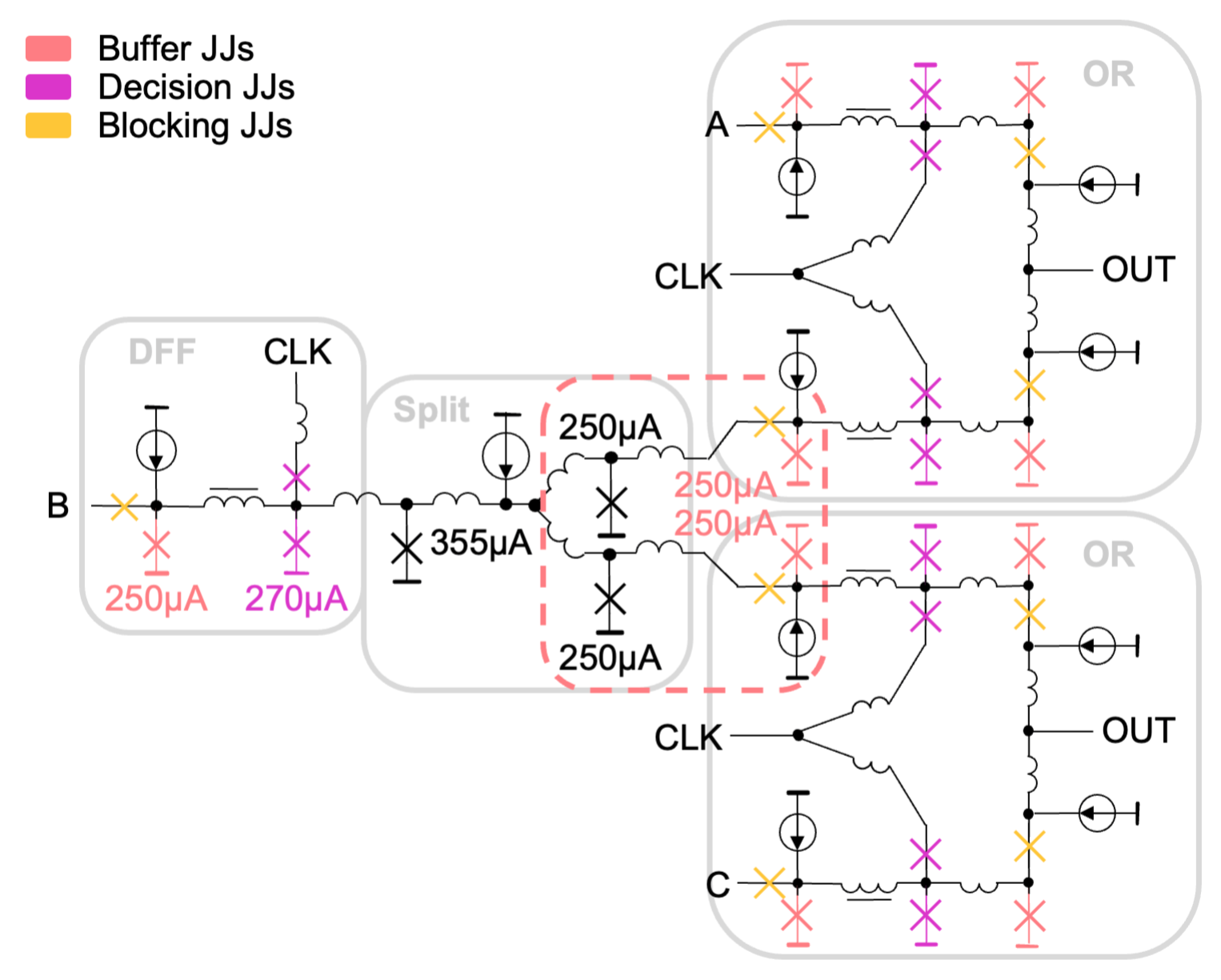}
    \caption{A signal splits from a D-flip flop (DFF) to the inputs of two OR cells. Cells are outlined and labeled with their names. A conventional splitter enables an FO2 from the DFF. The JJs at the outputs of the splitter are redundant, as they have the same critical currents as those on the inputs of the two OR cells and perform a similar buffering function. Thus, \revisedd{they can be dropped and }the OR cell's buffer JJs can be \revisedd{shared} for splitting.
    }
    \label{fig:dffsplitandmerge}
\end{figure}

\begin{figure}
    \centering
    \includegraphics[scale=0.13]{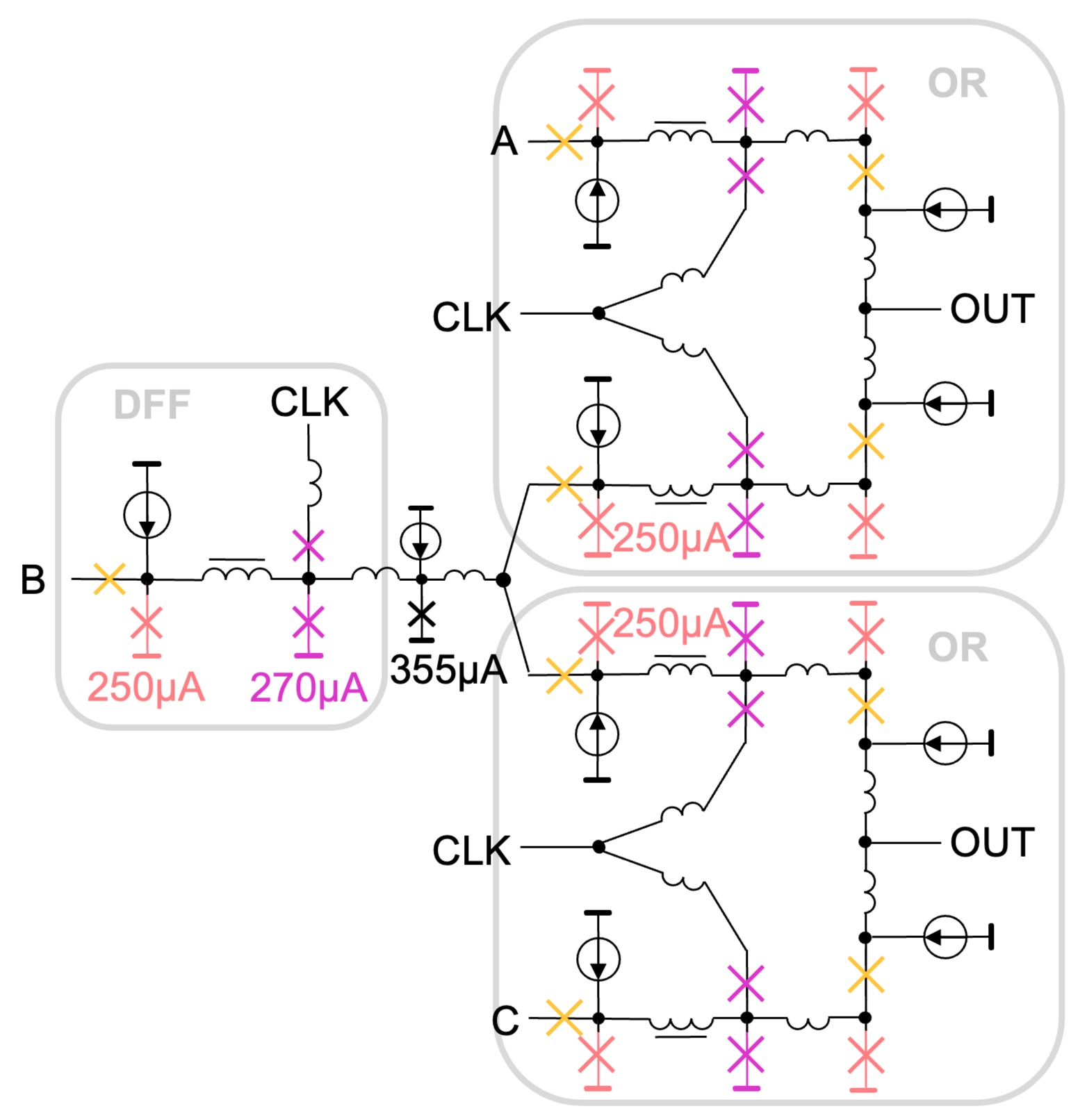}
    \caption{The consequences of \revisedd{sharing} the two OR cells' input JJs identified in Figure~\ref{fig:dffsplitandmerge}. The splitter's output JJs are ``merged'' with the OR cell's buffer JJs so that only one of these sets is used while the splitting task is still accomplished.
    }
    \label{fig:dffsplitandmerge2}
\end{figure}

\begin{figure}
    \centering
    \includegraphics[scale=0.13]{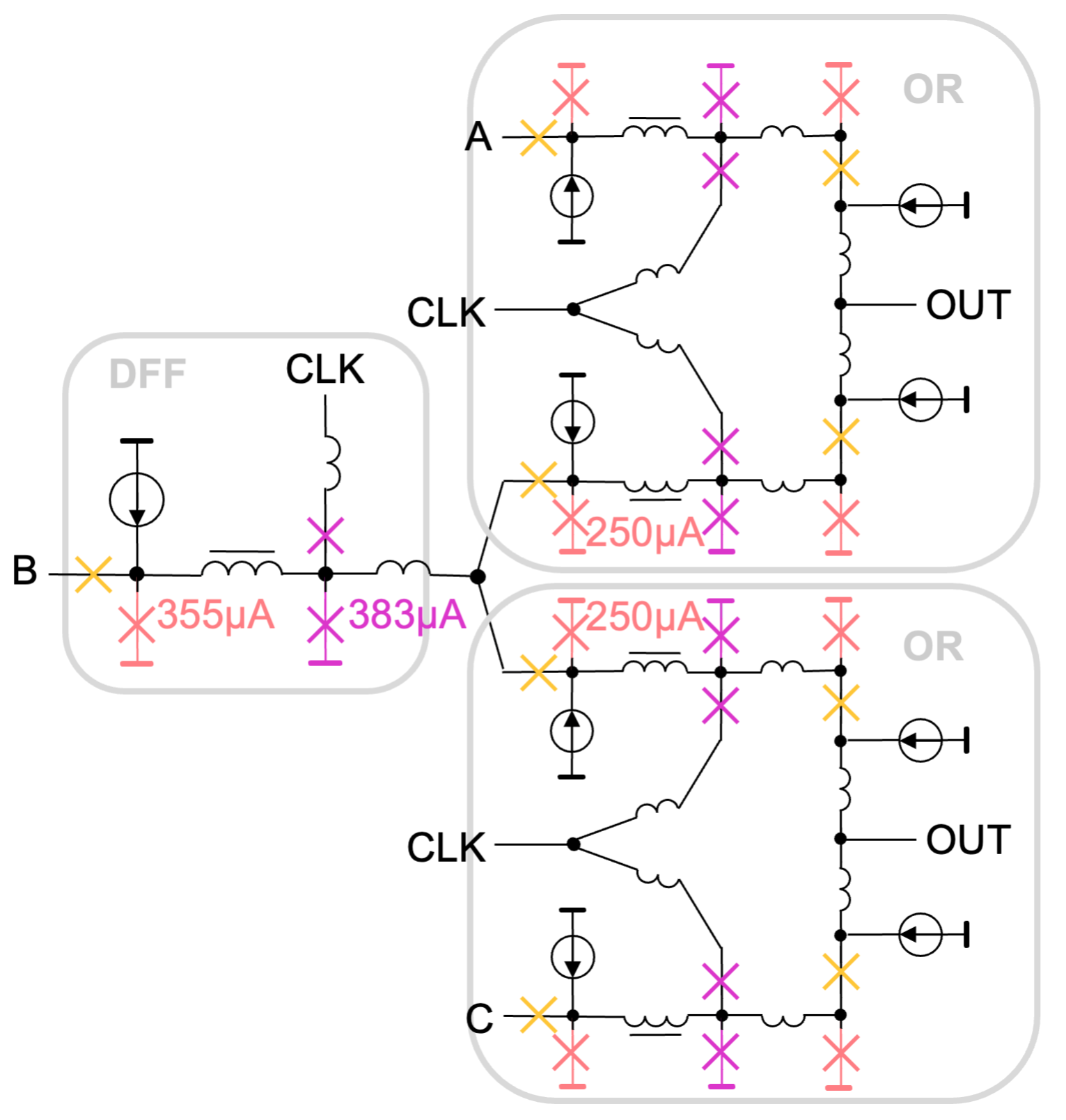}
    \caption{The 355~$\mu$A-JJ in Fig.~\ref{fig:dffsplitandmerge2}, left over from the splitter, is absorbed by the driving DFF through additional JJ \revisedd{sharing}. As a prerequisite, the DFF's components must be scaled so that the baseline critical current matches that of the absorbed JJ. The JJ that sets the baseline critical current is the \revised{pink} buffer JJ on the DFF input.
    }
    \label{fig:dffrank}
\end{figure}

To enable JJ \revisedd{sharing}, we look for redundancies in the logic cells. We start by examining the structure of conventional cells and categorizing JJs within them. In doing so, we pin down three common types: those used in decision-making pairs~\cite{fourieoverview18}, blocking JJs, and buffer JJs. An example OR cell~\cite{polonskyOR} is shown in \revised{Fig.}~\ref{fig:orgate} with all JJs color-coded based on their functions. Decision-making pairs (\revised{magenta}) act as comparators that consist of two JJs, in which one of the two JJs switches depending on the direction of bias current and incoming SFQ. Blocking JJs (\revised{gold}) reject the passage of SFQ pulses in two ways. If they exist at the inputs, they prevent an extra SFQ from entering a Superconducting Quantum Interference Device (SQUID) while one is already circulating; \revised{e.g., in the case of the leftmost \revised{gold} JJs in Fig.~\ref{fig:orgate}}. If they exist at the output, they inhibit the backward propagation of an SFQ towards the opposite input wire, as in the rightmost \revised{gold} JJs. Finally, buffer JJs (\revised{pink}) are inserted to maintain signal fidelity and typically serve to improve the electrical robustness of cells. In some cases, they also make up one side of a SQUID, with the decision-making JJ on the other side.

\subsection{JJ \revisedd{Sharing}}

The ideal JJs for \revisedd{sharing} sit on the edges of the cell and are wired in shunt configuration---as such, we call them ``boundary JJs''. In the examples that follow, both buffer JJs and decision-making pairs will be demonstrated for use in \revisedd{JJ sharing}. \revised{Fig.}~\ref{fig:dffsplitandmerge} shows a typical circuit consisting of a DFF, two OR cells, and a splitter, with a total JJ count of 31. \revised{Fig.}~\ref{fig:dffsplitandmerge2} shows the equivalent design after the proposed JJ \revisedd{sharing}. As can be seen, the logic cells effectively merge with the splitter, thereby saving 2 JJs.

\begin{figure*}
    \centering
    \includegraphics[scale=0.14]{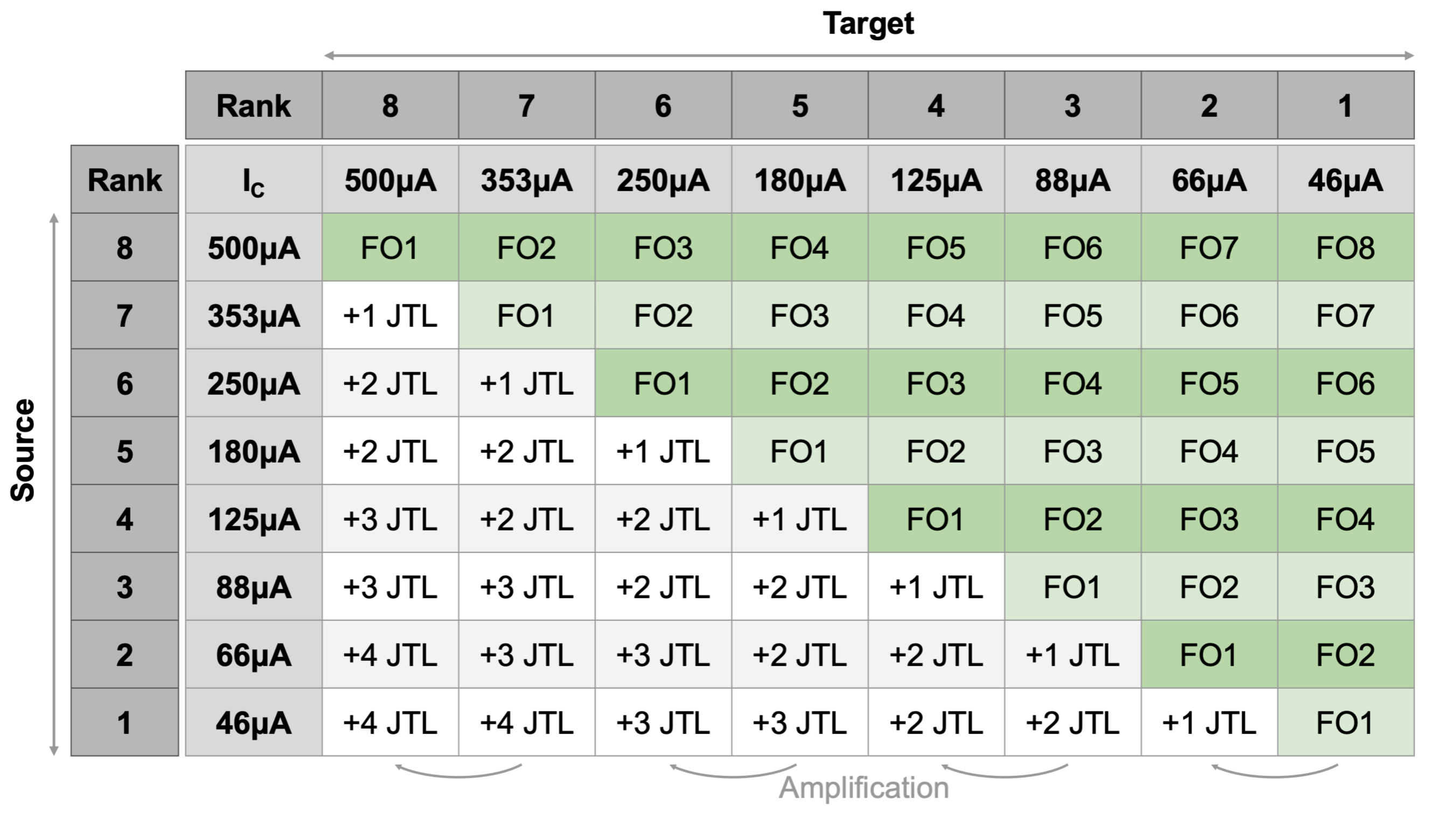}
    \caption{Rounded baseline critical current values for cells are abstracted into a property called ``rank''. This table depicts a rule set that dictates the connection type of a source cell (row) to the target cell(s) (column). The FO1 diagonal between green and gray/white cells marks a transitional line---to the right, the connection is \textit{fan-out positive} and needs no additional hardware to achieve a fan-out up to the amount shown; to the left, the connection is fan-out negative: each cell lists the number of amplifying JTLs needed to bridge the gap in rank. The hardware overhead required to amplify from a lower rank to a higher one increases sub-linearly with the distance between ranks, as one additional step in rank can be gained \textit{between} amplifying JTLs. \revisedd{Each JTL comprises two JJs.}
    }
    \label{fig:fanoutcost}
\end{figure*}

The above reassignment is made possible by the fact that the \revisedd{shared} JJs have the same sizes before \revisedd{sharing}. However, one 355~$\mu$A-JJ (in black) remains in \revised{Fig.}~\ref{fig:dffsplitandmerge2}, left over by the splitter. Convention states that because the baseline critical current of every cell in a library is static---for example, 250~$\mu$A~\cite{likharev_semenov91,coldfluxlib}---this JJ cannot be absorbed by the DFF that precedes it. Instead, we consider here a case where the baseline current can be flexibly assigned. For instance, if the DFF's decision-making pair I$_\text{C}$ is scaled by 355~$\mu$A to come close to the I$_\text{C}$ of the splitter's input JJ, then it can be \revisedd{shared} for splitting. The result is shown in \revised{Fig.}~\ref{fig:dffrank}, in which all other JJs in the DFF are scaled by the same number\footnote{\reviseddd{In practice, all cell inductors and resistors are inversely scaled by this value.}}. \revised{By sharing from both sides}, all of the JJs in the original splitter \revised{of \revised{Fig.}~\ref{fig:dffsplitandmerge}} are absorbed to create a savings of three JJs, and splitting occurs directly at the output of the modified DFF. \revised{This modification's effect on bias margins depends on cell optimizations with different fan-out requirements and on the critical current gap between source and target cells. Cell-specific optimizations are beyond the scope of this paper, but basic design rules that guide more favorable connections are organized and presented in the next section.}

\section{I$_\text{C}$ Abstraction Through Ranking}\label{methods}

Even in small fan-out cases, critical current choices are important to ensure reliable connectivity. For example, an FO2 connection requires all of the JJs in the driving cell to be scaled by $\sqrt{2}$~\cite{likharev_semenov91}. Applying this approach to cells with larger fan-outs and baseline critical currents that are potentially different from their neighbors', however, turns every fan-out point into a unique calculation, which complicates the design process. 

To simplify the procedure for connecting these cells, we create an abstraction in which critical currents are discrete. \revised{We call this abstraction ranking}, wherein we select specific current values that have consistent separation within a range. 
Through ranking, we constrain the design options in a way that enforces the current ratio requirements of a connection. In this section, we define the relationship between the critical current of the driving boundary JJ, the critical current(s) of the target boundary JJ(s), and the maximum fan-out. We then detail the cost of various connections \revised{in terms of JJ count} and finally compare and contrast three different rank-based design methodologies. 

\subsection{Ranking}~\label{sec:ranking}
A cell's baseline critical current defines its maximum fan-out capacity, and therefore it is considered the primary identifying electrical characteristic. To relate critical current, which is analog, to fan-out capacity, which is discrete, we define the translation from one domain to the other.

Nominal values are chosen to start from 250~$\mu$A, as it is a popular baseline critical current for SFQ libraries~\cite{coldfluxlib,likharev_semenov91} and used in the prior example in \revised{Fig.}~\ref{fig:dffsplitandmerge} and \revised{Fig.~\ref{fig:dffsplitandmerge2}}. To discretize the range of critical currents, we take this as a central point and explore steps below and above it by multiplying by $(1/\sqrt{2})$ and $\sqrt{2}$, respectively. Going down, we reach 180~$\mu$A, 125~$\mu$A, 88~$\mu$A, and 66~$\mu$A in turn, and finally, stop at 46~$\mu$A. Below this point, JJs become more prone to thermal errors~\cite{tolpygo2016superconductor,fabprocess15}. Applying the same approach to the other side, 353~$\mu$A is found after one step and 500~$\mu$A after two, which we choose here as our upper bound. \revised{In practice, however, the maximum JJ size is limited by the Josephson penetration depth~\cite{doi:10.1063/1.5124391} and practical shunt resistor size.} \revisedd{Additionally, it should be noted that the proposed JJ sharing and cell ranking techniques are independent of any particular selection of critical currents, range, and step size. }\revisedd{From here on, we abstract the eight above-mentioned cell baseline critical currents with number labels, where the lowest baseline is assigned a label of 1 and the highest a label of 8.}

\revised{In general, the number of ranks between the minimum and maximum critical currents with a rank step size $p_r$ is
\begin{equation}
    N_r \ge \frac{\log (\frac{I_{C,Max}}{I_{C,Min}})}{\log (p_r)} + 1
\end{equation} 
while the number of JTLs needed to amplify from a source critical current I$_\text{S}$ to a target current I$_\text{T}$ with an amplification step size $p_a$ is 
\begin{equation}
    N_{JTL} \ge \frac{\log (\frac{I_T}{I_S})}{2 \log (p_a)}.
\end{equation}
}\revised{The above inequalities stem from the fact that $N_r, N_{JTL} \in \mathbb{N}$ and the observation that not every rank's nominal critical current value will necessarily match those chosen in an existing cell design.} To add robustness and maximize interoperability between different library cells that may not match perfectly, critical current intervals can be assigned; however, such an assignment is beyond the scope of this paper.

\revised{Fig.}~\ref{fig:fanoutcost} shows \revised{a lookup table that describes} the connectivity between various ranks for the range discussed above \revised{and $p_r=p_a\approx\sqrt2$} . A cell with a higher rank, or larger baseline critical current, can be directly connected to any other with \reviseddd{the same or} lower rank. Consider an example in which four target cells' ranks must be found for a design that has \revised{an} FO4 from a rank-6 source cell with a 250~$\mu$A baseline critical current. It can be concluded from the crosspoint of a rank-6 source cell and FO4 cell value in \revised{Fig.}~\ref{fig:fanoutcost} that the target cells must be of rank-3.

Conversely, connecting a smaller-ranked cell to one with a higher rank necessitates the use of amplifying JTLs, the number of which is also dependent on their rank difference. Assume that it is now desirable to move one of those rank-3 cells back up to rank-6, likely in advance of a larger anticipated fan-out. In this case, the source is a rank-3 cell and the target is rank-6. The crosspoint of the rank-3 row and the rank-6 column in \revised{Fig.}~\ref{fig:fanoutcost} indicate that two amplifying JTLs are needed for the connection. The first of the two JTLs will amplify from rank-3 to rank-4 and the second from rank-5 to rank-6. Note that an additional step in rank \revised{is} gained (rank-4 to rank-5) in-between JTLs without introducing another amplifying JTL.

\begin{figure}
    \centering
    \includegraphics[scale=0.115]{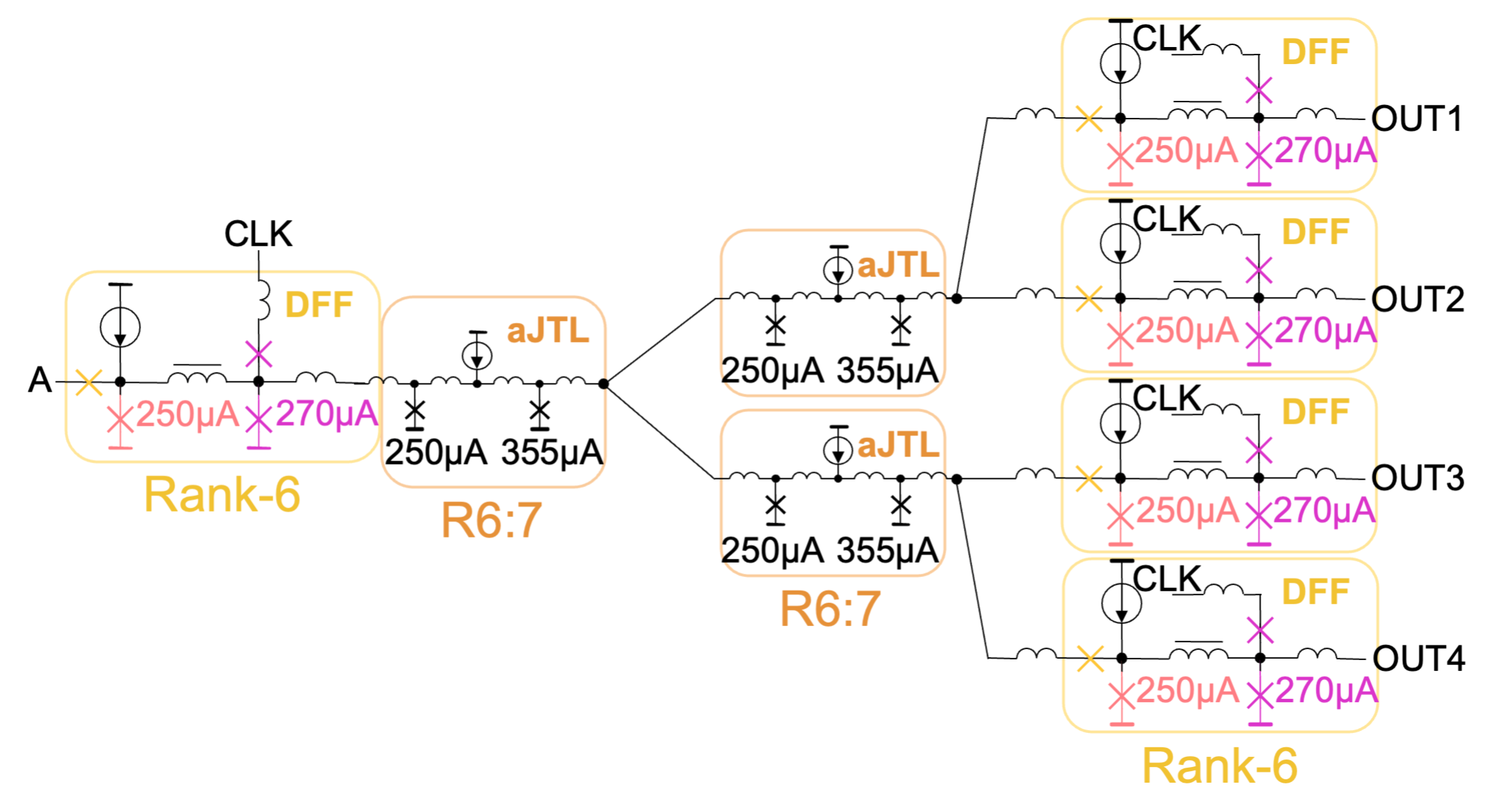}
    \caption{Ranking affords three primary design methodologies, each of which is demonstrated with an example. A single DFF source drives four DFF targets. Cells are outlined and labeled with their names and ranks. Amplifying JTLs modify the rank from input to output where needed, and are labeled as \revised{R$s$:$t$, where $s$ and $t$ are the source and target} ranks, respectively. The nominal rank \reviseddd{is chosen to match} the baseline used in the original library; here, it is 250~$\mu$A. Fan-out is achieved by using amplifying JTLs\revised{---in this case, they are arranged in a splitter-like tree structure}.
    }
    \label{fig:designmeth1}
\end{figure}

\begin{figure}
    \centering
    \includegraphics[scale=0.12]{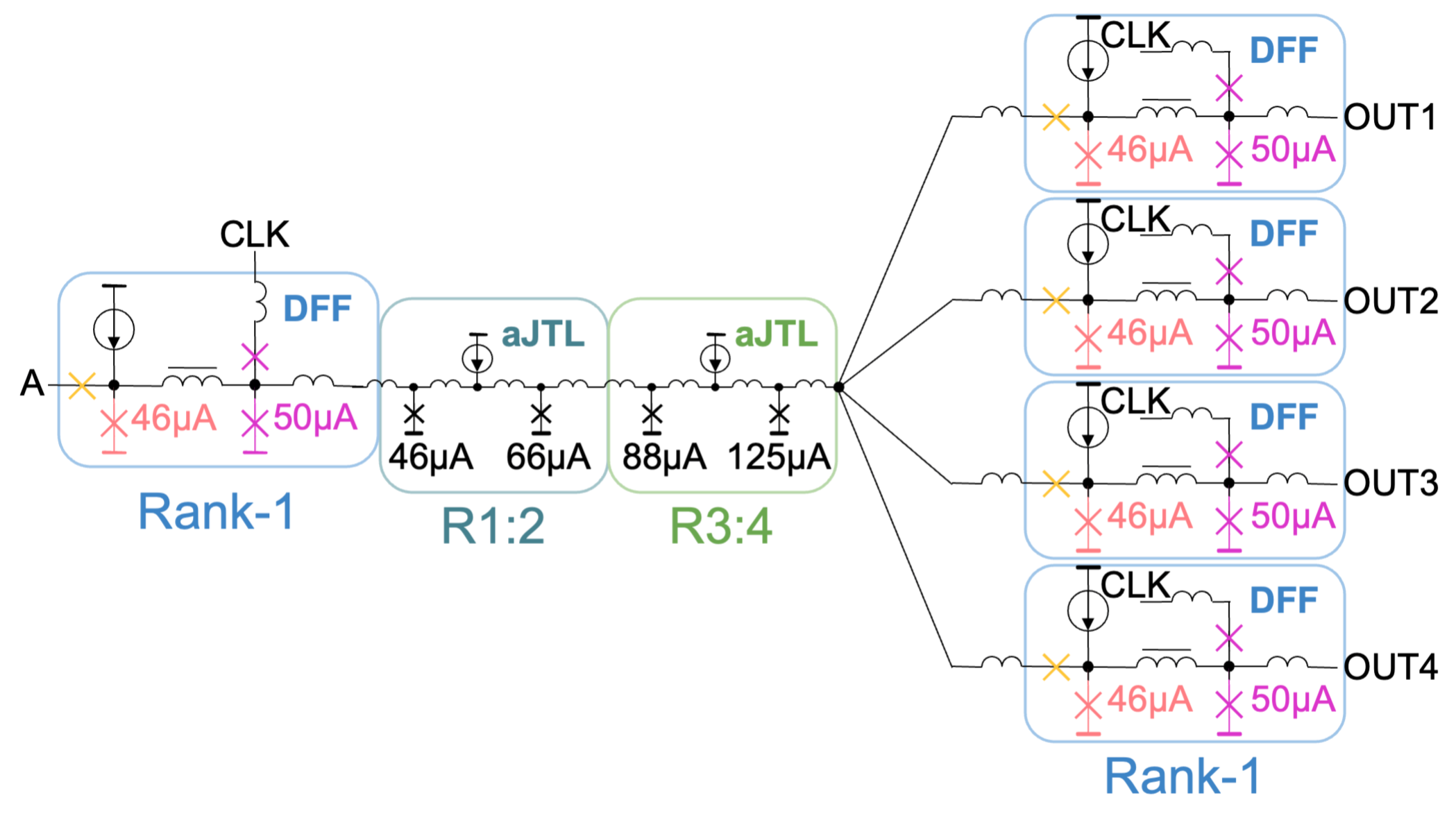}
    \caption{The second rank-based design methodology. The baseline rank is 1 and is amplified only in advance of fan-out.
    }
    \label{fig:designmeth2}
\end{figure}

\begin{figure}
    \centering
    \includegraphics[scale=0.12]{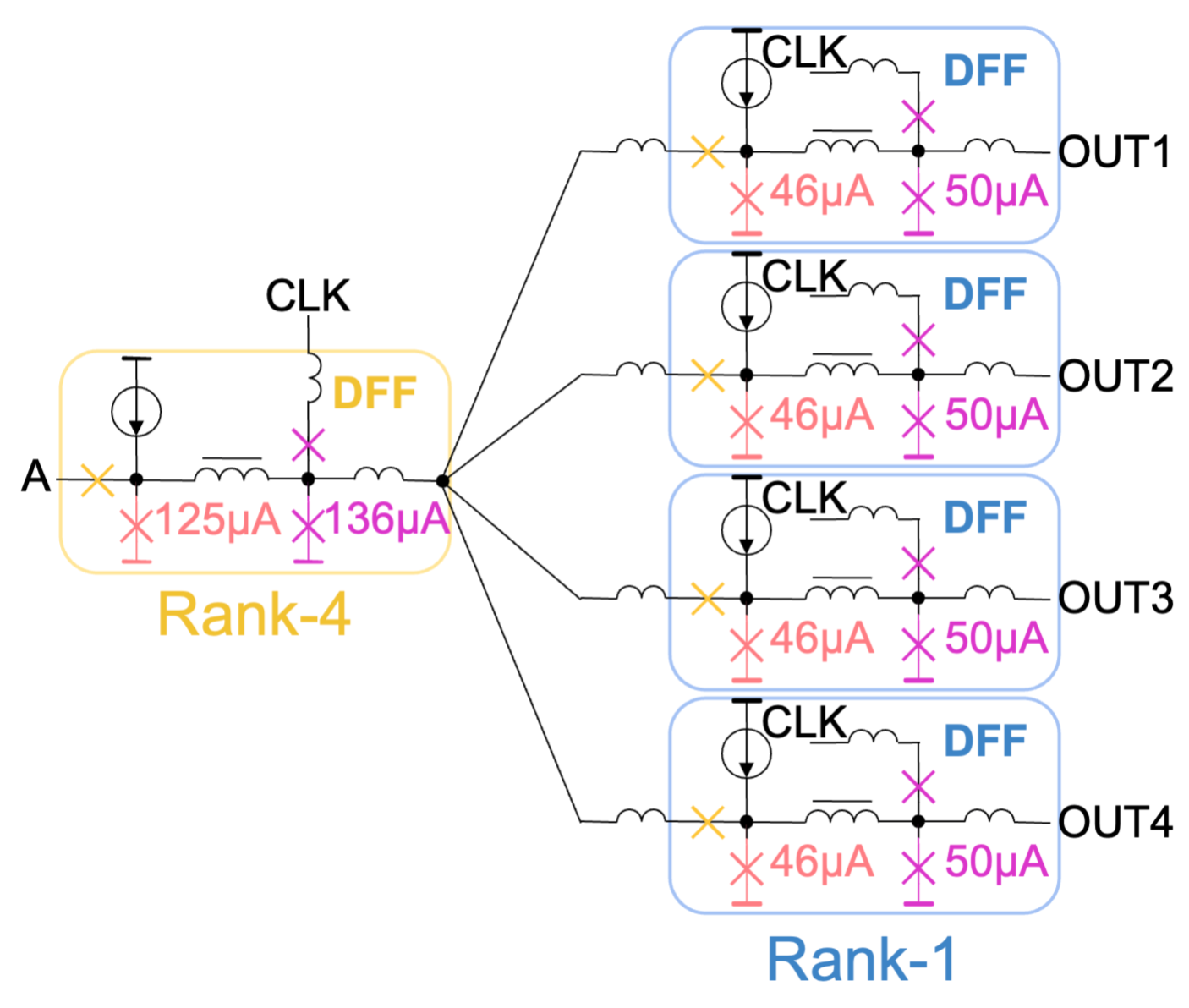}
    \caption{The third rank-based design methodology is a flexible ranking scheme, in which the source cell and target cell(s) can take on any rank as long as the fan-out rules in \revised{Fig.}~\ref{fig:fanoutcost} are upheld.
    }
    \label{fig:designmeth3}
\end{figure}

\subsection{Design Methods}

The rules discussed above can be \reviseddd{carried out} in more than one way depending on the designer's target goals. Below, we use a DFF that fans out to four more DFFs as a running example \reviseddd{to demonstrate three such design methods. In Section~\ref{tradeoffs}, we elaborate on them} and show how one can apply ranking to optimize for design reuse, energy, \revisedd{speed, or} JJ count.

The first method requires the rank to remain the same for all logic cells. This enables the use of logic cells from existing libraries~\cite{likharev_semenov91,coldfluxlib} without modifications. An example is provided in \revised{Fig.}~\ref{fig:designmeth1}, where cells of rank-6 make up the baseline. To achieve an FO4, \revised{three amplifying JTLs can be chained together in a tree structure, much like conventional splitters. Each JTL amplifies by one rank and fans out to two lines, saving three JJs compared to the conventional case.}

The second method keeps the logic cells at the lowest rank, which provides faster JJ switching speeds and lower power consumption~\cite{tolpygo2016superconductor,ultimateperf2}. Similar to the first case, the logic cell library needs no modifications after its initial design because the rank is consistent---only one copy of each cell is needed. However, because the starting rank is the lowest, it is likely to save more JJs at higher fan-outs compared to the first case. \revised{Fig.}~\ref{fig:designmeth2} depicts the resulting schematic. In this case, a two-stage JTL chain is needed before fan-out. The JJ count is five fewer than that using the splitter method.

The third method aims to reduce the number of amplifying JTLs used in the previous methods. Conventionally, superconductor cell libraries select a single baseline critical current to be used globally for all cells. However, adopting a more flexible assignment allows the fan-out point to shift closer to the output of the source logic cell. Applying this to the running example results in the circuit shown in \revised{Fig.}~\ref{fig:designmeth3}, which begins with a rank-4 DFF that immediately splits to rank-1 DFFs \reviseddd{and thus, in this case, removes all hardware overhead for splitting.}

\section{Evaluation}
\label{eval}
\subsection{Evaluating the Basic Building Blocks}
\label{sec:basicbuildingblox}

\subsubsection{Bias Margin Analysis}

The JTL chain is a fundamental building block of SFQ design and critical for the proposed approach. For this reason, various \revised{amplification} chain configurations are simulated extensively\revised{, similarly to splitter chains,} in the following experiments and their bias margins are compared. \revisedd{All designs are simulated in the Cadence design suite using models based on the MIT Lincoln Laboratory SFQ5ee 10 $kA/cm^2$ process \reviseddd{and are terminated with strings of non-amplifying JTLs that match the designs' output impedances.}} Each JTL element consists of two \revisedd{externally-shunted} JJs, four inductors, and a single resistor that biases both JJs. Each JTL chain that amplifies the rank from \revised{R$s$ to R$t$} is labeled with \revised{R$s$:$t$}. For a JTL that is not amplifying, \revised{$s=t$}.

\begin{figure} [ht]
    \centering
    \includegraphics[scale=0.47]{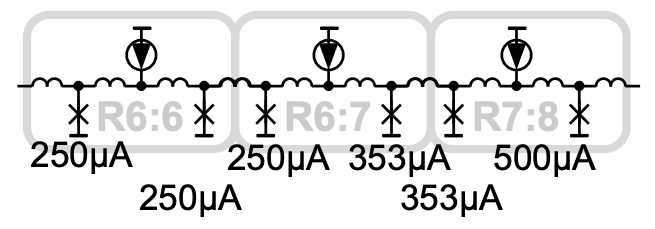}
    \caption{Amplifying JTL chain\revisedd{---the first JTL stage serves as a non-amplifying buffer between test inputs and the circuit under test}. Bias margins are \revisedd{simulated and found} to be \reviseddd{$+38.5\%/-50\%$}.}
    \label{fig:jtltest1}
\end{figure}

\revisedd{Our starting case is a} JTL chain without amplification, designed with rank-6 JTLs. Bias margin \revisedd{simulations} indicate an operating range of \revised{$+38.5\%/-65.4\%$}. A chain that amplifies from a rank of six to a rank of eight with a step size of $\sqrt{2}$~\cite{likharev_semenov91,wilkinson} is depicted in \revised{Fig.}~\ref{fig:jtltest1} and also has margins of \reviseddd{$+38.5\%/-50\%$.} \revisedd{The first JTL stage acts as a non-amplifying buffer between test inputs and the circuit under test}. These two \revised{results} serve as a reference for the ensuing tests.

\begin{figure}[ht]
    \centering
    \includegraphics[scale=0.47]{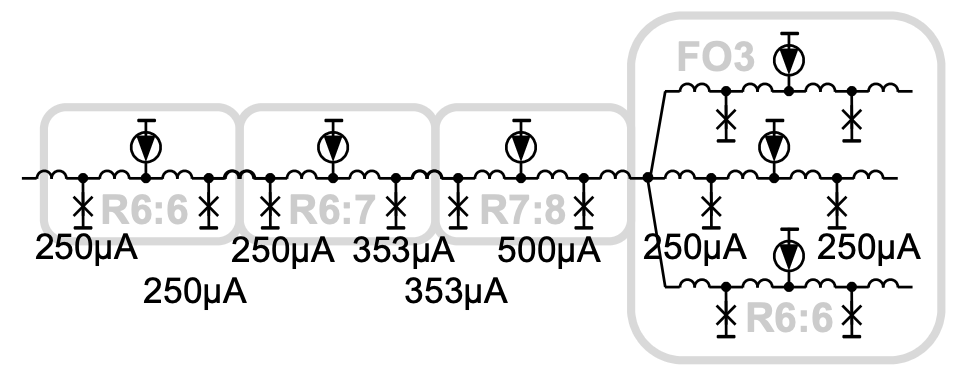}
    \caption{Amplifying JTL chain leading to FO3\revisedd{---the first JTL stage serves as a non-amplifying buffer between test inputs and the circuit under test}. Bias margins are \revisedd{simulated and found} to be $+38.5\%/-38.5\%$.}
    \label{fig:jtltest2}
\end{figure}

To test fan-out directly from the JTLs, the above chain is used to split to three chains of rank-6 JTLs. This configuration is shown in \revised{Fig.}~\ref{fig:jtltest2} and has bias margins of \reviseddd{$+38.5\%/-38.5\%$}. \revised{By comparison, a splitter tree with the same fan-out, composed of directly-chained splitters that match Fig.~\ref{fig:splitter} and have a current I$_\text{X}$ of 250 $\mu$A, has bias margins of \reviseddd{$+26.9\%/-30.8\%$}.} 

\begin{figure}[ht]
    \centering
    \includegraphics[scale=0.47]{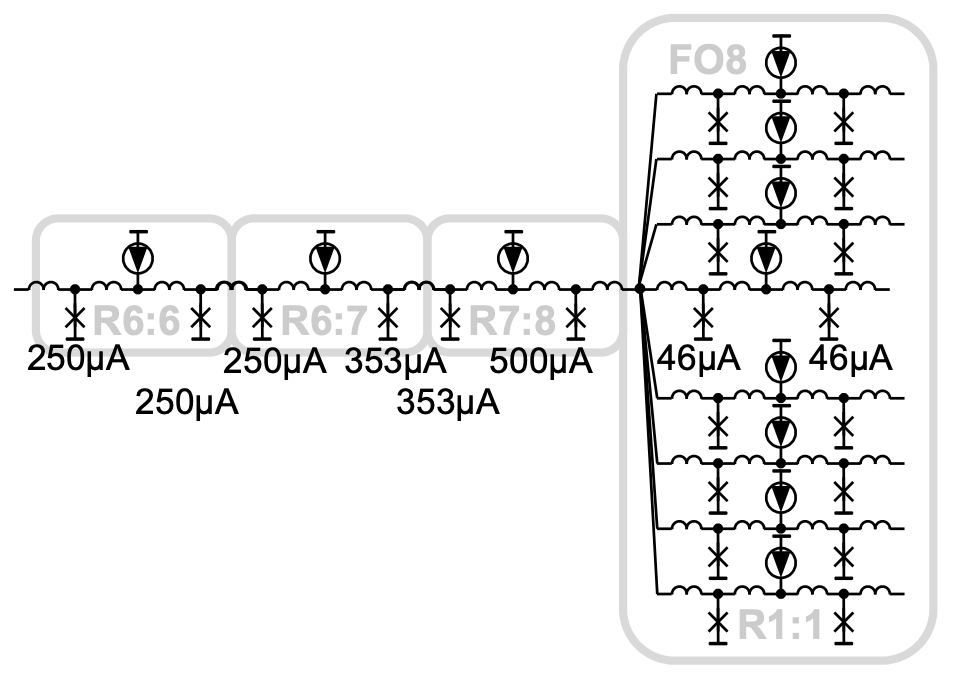}
    \caption{Amplifying JTL chain leading to FO8\revisedd{---the first JTL stage serves as a non-amplifying buffer between test inputs and the circuit under test}. Bias margins are \revisedd{simulated and found} to be $+38.5\%/-53.8\%$.}
    \label{fig:jtltest3}
\end{figure}

To test the effects of additional fan-out, five more JTL chains are added onto the fan-out node and all of the output chains' baselines are reduced to rank-1 to achieve an FO8 in total. The schematic is shown in \revised{Fig.}~\ref{fig:jtltest3}. The bias margins, in this case, are \reviseddd{$+38.5\%/-53.8\%$}. \revised{By comparison, a splitter tree with an FO8, composed of the same splitters described above, has bias margins of \reviseddd{$+26.9\%/-26.9\%$}.} 

Next, amplification is added from rank-1 to rank-8 to lead into the FO8, using a step size of $\sqrt{2}$ within the JTLs and between JTLs, yielding bias margins of \reviseddd{$+42.3\%/-23.1\%$}. This serves as a \revisedd{simulation} of the longest possible amplification chain with the given rank range.

\begin{figure}[ht]
    \centering
    \includegraphics[scale=0.18]{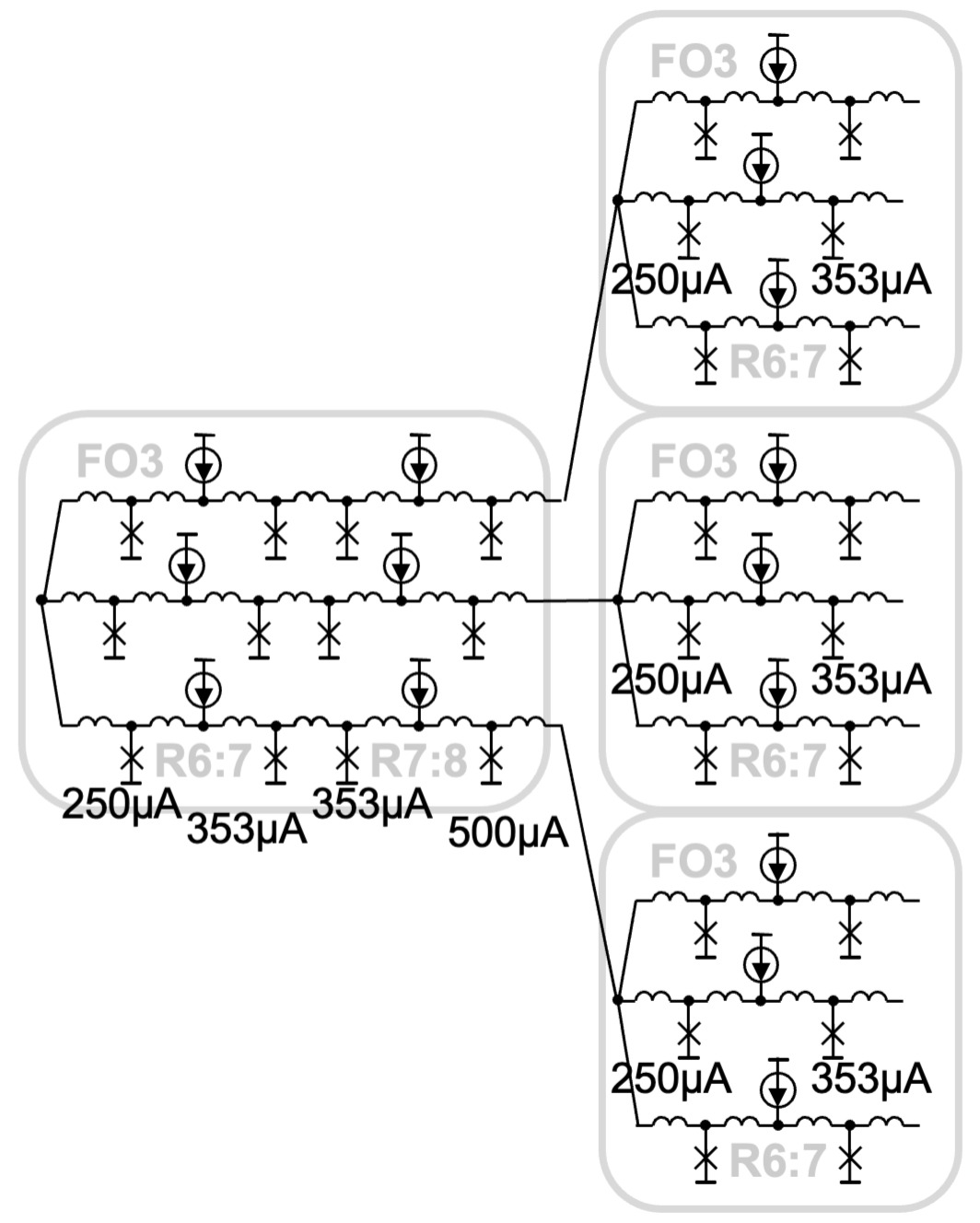}
    \caption{Amplifying JTL chain leading to FO9, using R6:8 amplifying segments. Bias margins are \revisedd{simulated and found} to be \reviseddd{$+38.5\%/-30.8\%$}.}
    \label{fig:jtltest4}
\end{figure}

The above experiments show that bias margins are, for the most part, maintained after each progression, starting from the FO3 in \revised{Fig.}~\ref{fig:jtltest2}. For the sake of modularizing the design, it is important to know whether it is possible to ultimately achieve the same fan-out with different cells while preserving those cells' respective bias margins. We test FO9 using the same JTL chains that amplify from rank-6 to rank-8, as shown in \revised{Fig.}~\ref{fig:jtltest4}, and find that bias margins are \reviseddd{$+38.5\%/-30.8\%$}, which are similar to those of the FO3 case above and reveal that the design method indeed promotes modularity.

~\revised{Bias margin \revisedd{simulation results} for the above examples are summarized in Fig.~\ref{marginsummary}}.

\begin{figure} [ht]
\begin{center}
\resizebox{\columnwidth}{!}{%
\reviseddd{
\begin{tabular}{|l|c|c|} 
\hline
\textbf{Configuration} & \textbf{Ranking} & \textbf{Splitters}  \\ 
\hline
\textbf{R6:8} (Fig.~\ref{fig:jtltest1})           & $+38.5\%/-50\%$      & -       \\ 
\hline
\textbf{R6:8+FO3} (Fig.~\ref{fig:jtltest2})          & $+38.5\%/-38.5\%$      & $+26.9\%/-30.8\%$     \\ 
\hline
\textbf{R6:8+FO8} (Fig.~\ref{fig:jtltest3})          & $+38.5\%/-53.8\%$      & $+26.9\%/-26.9\%$     \\ 
\hline
\textbf{R1:8+FO8}           & $+42.3\%/-23.1\%$       & $+26.9\%/-26.9\%$                   \\
\hline
\textbf{R6:8+FO9} (Fig.~\ref{fig:jtltest4})           & $+38.5\%/-30.8\%$       & $+30.8\%/-26.9\%$                   \\
\hline
\end{tabular}}
}
\caption{\revised{Summary of bias margins and comparison of various fan-out configurations between conventional splitters and ranking. The fourth entry refers to the configuration in which an amplifying JTL chain from rank-1 to rank-8 precedes a fan-out to 8.}}\label{marginsummary}
\end{center}
\end{figure}

\subsubsection{Current Savings Analysis}
\revised{
According to the above-presented analyses, the proposed rank-based design methodology shows promise for better resource utilization without bias margin degradation. However, it is not clear that bias current is conserved, as different ranks impose different bias current requirements. To shed light on this, we calculate the bias current overhead for fanning out with an FO2, FO4, and FO8 from various logic cells and compare the ranking methodology to conventional splitting \reviseddd{in terms of bias current savings}. Our results are shown in Fig. \ref{tab:biasestimates}. In this study, we consider two cases: in the first, we use flexible ranking, in which the source cell is of rank-6 and the target cells assume any rank that uses the smallest bias current overhead. In the second option, the source and target cell ranks match.}

\begin{figure}
\begin{centering}
\revised{
\resizebox{0.78\columnwidth}{!}{%
\begin{tabular}{|c|c|c|c|} 
\hline
\textbf{Cell} & \textbf{Fan-Out} & \textbf{Flex. Rank} & \textbf{Matched Rank}  \\ 
\hline
\textbf{AND}  & \textit{FO2}     & 59.2\%             & 17.4\%                \\ 
\hline
              & \textit{FO4}     & 81.3\%             & 23.9\%                \\ 
\hline
              & \textit{FO8}     & 68.9\%             & 26.8\%                \\ 
\hline
\textbf{OR}   & \textit{FO2}     & 49.3\%             & 14.5\%                \\ 
\hline
              & \textit{FO4}     & 74.4\%             & 21.9\%                \\ 
\hline
              & \textit{FO8}     & 66.0\%             & 25.6\%                \\ 
\hline
\textbf{XOR}  & \textit{FO2}     & 77.6\%             & 22.8\%                \\ 
\hline
              & \textit{FO4}     & 91.2\%             & 26.8\%                \\ 
\hline
              & \textit{FO8}     & 72.7\%             & 28.3\%                \\ 
\hline
\textbf{INV}  & \textit{FO2}     & 41.8\%             & 12.3\%                \\ 
\hline
              & \textit{FO4}     & 68.3\%             & 20.1\%                \\ 
\hline
              & \textit{FO8}     & 63.1\%             & 24.5\%                \\ 
\hline
\textbf{DFF}  & \textit{FO2}     & 77.6\%             & 22.8\%                \\ 
\hline
              & \textit{FO4}     & 91.2\%             & 26.8\%                \\ 
\hline
              & \textit{FO8}     & 72.7\%             & 28.3\%                \\ 
\hline
\textbf{AVG}  &                  & 70.4\%             & 22.9\%                \\
\hline
\end{tabular}
}
\caption{\revised{Percent savings in total bias current of various basic cell configurations with ranking versus conventional splitting. Two ranking options are considered: flexible ranking, in which the source cell is of rank-6 and the target cells take on any desired rank; and matched ranking, in which both the source and target cells are of rank-6.
}}\label{tab:biasestimates}}
\end{centering}
\end{figure}

\begin{figure*}[ht]
    \centering
    \includegraphics[scale=0.16]{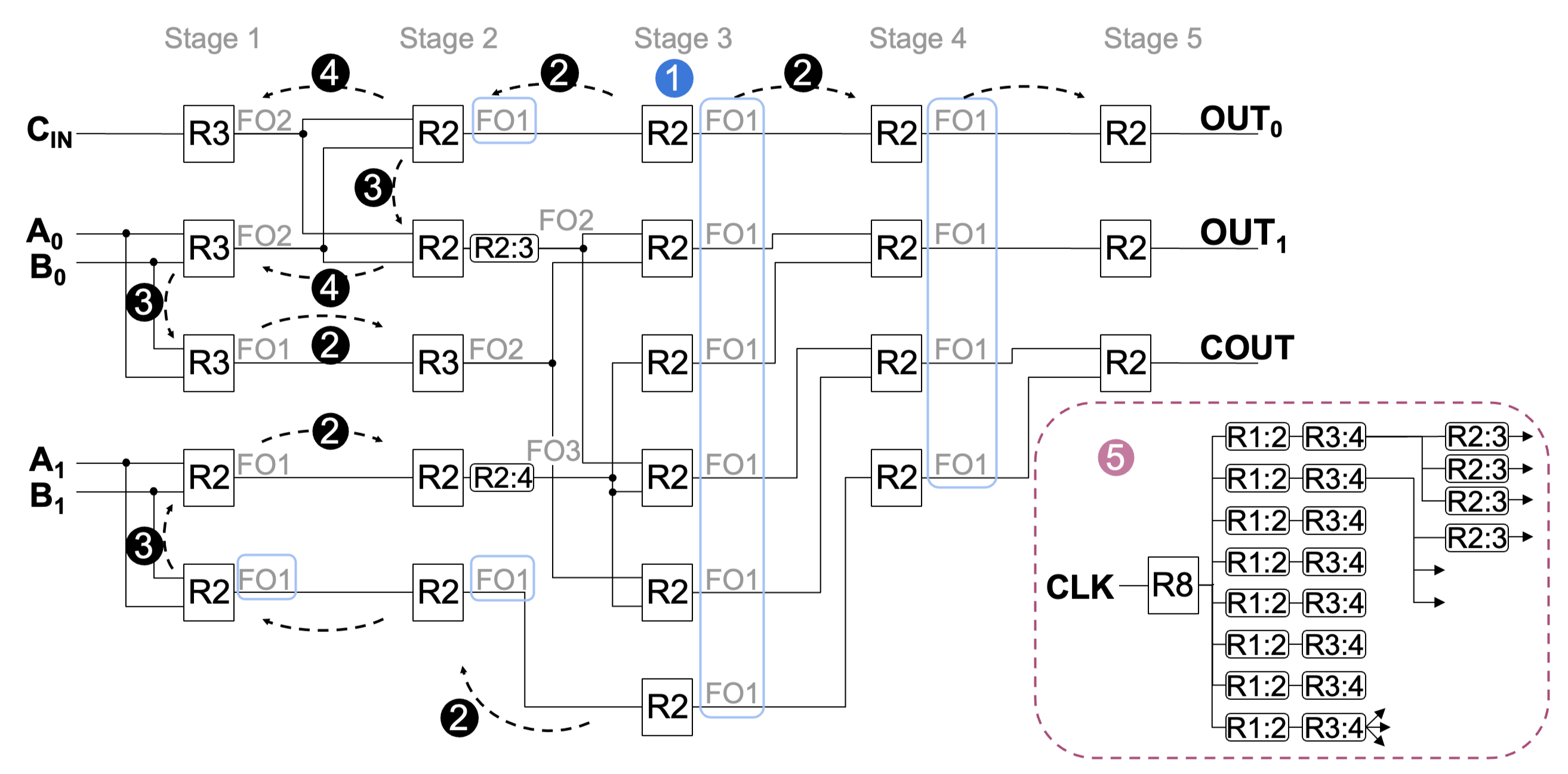}
    \caption{Block diagram of a fully-synchronous 2-bit Kogge Stone Adder (KSA) designed and simulated in Cadence using the MITLL SFQ5ee 100~$\mu$A/$\mu$m$^2$ fabrication process models. Ranks are assigned to the design and clock tree using a five-step methodology. The proposed \revisedd{JJ sharing} and flexible $I_C$ assignment save 17.7\% of JJs compared to the conventional splitting method.
    }
    \label{fig:ksasch}
\end{figure*}

\begin{figure}
    \centering
    \includegraphics[scale=0.35]{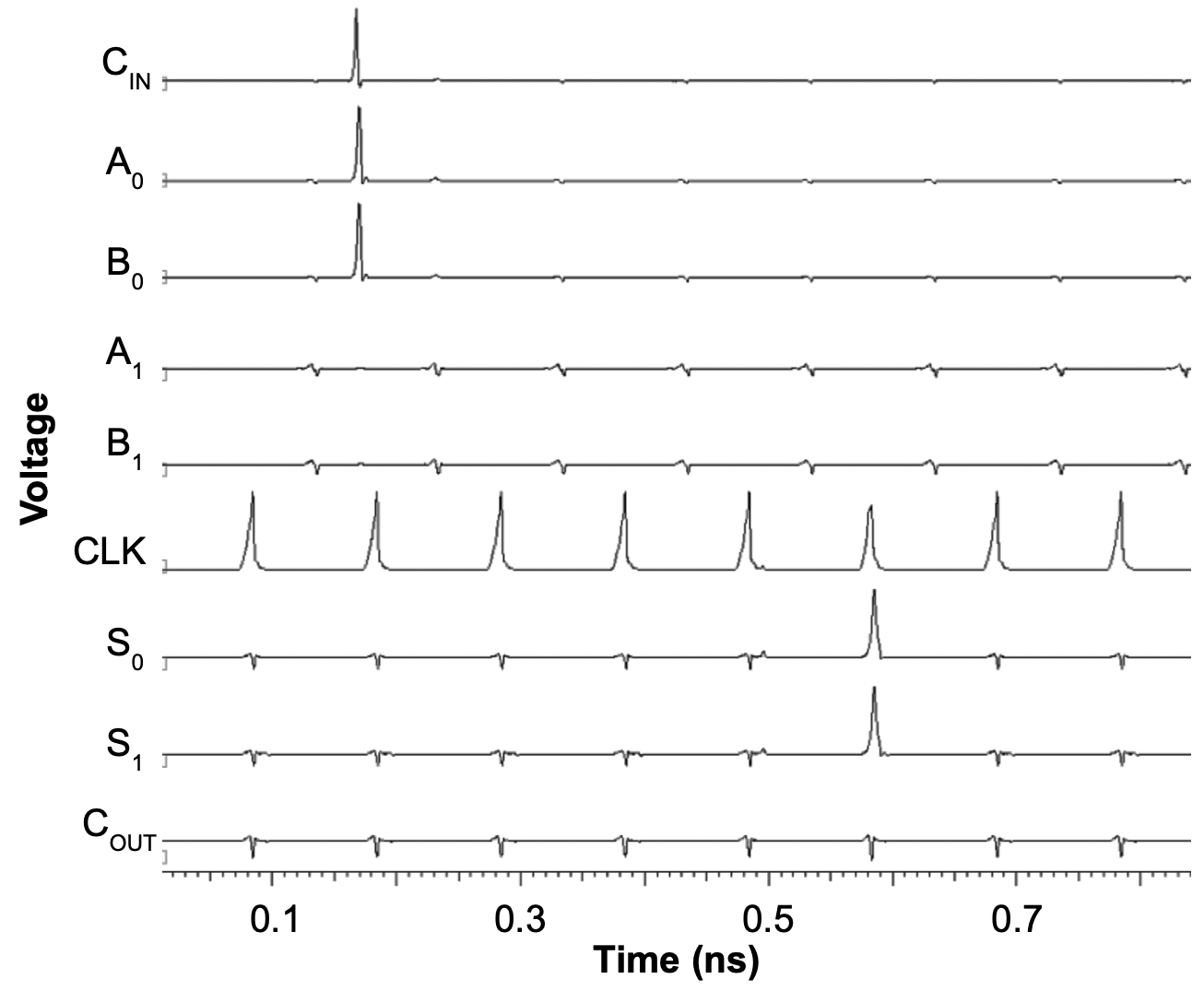}
    \caption{Simulation waveform for the 2-bit KSA with ranking. Inputs are 01, 01, and 1 for $A$, $B$, and $C_{IN}$, respectively. The correct result, $C_{OUT}=0$, $S_1=1$, and $S_0=1$, appears five clock cycles later.}
    \label{fig:ksasim}
\end{figure}

\subsection{Building Rank-Based Circuits}
Next, a 2-bit KSA is used as a comprehensible example to demonstrate how ranking can be used to simplify the design of circuits with flexible baseline I$_\text{C}$s. The block diagram of the adder design is shown in \revised{Fig.}~\ref{fig:ksasch}. Synchronous AND, OR, XOR, and DFF cells are used for its implementation. Considering that the logic function of each cell is not relevant to ranking, however, cells are depicted as rectangles only labeled with their ranks, \revised{R$s$}. As before, JTL chains that amplify the rank from \revised{R$s$ to R$t$}, necessary to meet the ranking rules from \revised{Fig.}~\ref{fig:fanoutcost}, are labeled as \revised{R$s$:$t$}. The required fan-out at the output of each cell is also labeled.

Assigning custom ranks to the fully-synchronous KSA design takes just five steps:
\begin{enumerate}
    \item \textbf{Step} \textcolor[rgb]{0.24,0.47,0.85}{\ding{202}}. Find the stage with the most synchronous elements and assign the highest or lowest rank needed to minimize the number of additional JTLs on the clock line. \revisedd{Ascribe the highest rank if additional fan-out is needed after this stage, or if lower variability is needed~\cite{tolpygo2016superconductor}; ascribe the lowest rank if greater energy-efficiency or speed is desired.} In the design of \revised{Fig.}~\ref{fig:ksasch}, the third stage has 6 synchronous components---more than any other stage. FO7 is possible with a clock that splits from rank-8 to rank-2 cells, or from rank-7 to rank-1 cells. We opt for the first case and assign rank-2 (R2) to all cells in this stage.
    
    \item \textbf{Step} \ding{203}. For any two cells that share a direct connection, FO1, assign the same rank to both. That is to say, chained FO1 connections will propagate the same rank value. In \revised{Fig.}~\ref{fig:ksasch}, we start at the stage found in \textcolor[rgb]{0.24,0.47,0.85}{\ding{202}} and propagate the values over all FO1 connections to the left and right sides. As a consequence, all cells in the fourth and fifth stages in \revised{Fig.}~\ref{fig:ksasch} are assigned rank-2. Additionally, the top cell in the second stage, and the bottom cells in the first and second stages, are assigned rank-2. Other cells that share an FO1 connection are noted to have the same ranks as their connected neighbors, which are not yet known.
    
    \begin{figure*}[ht]
    \centering
    \includegraphics[scale=0.36]{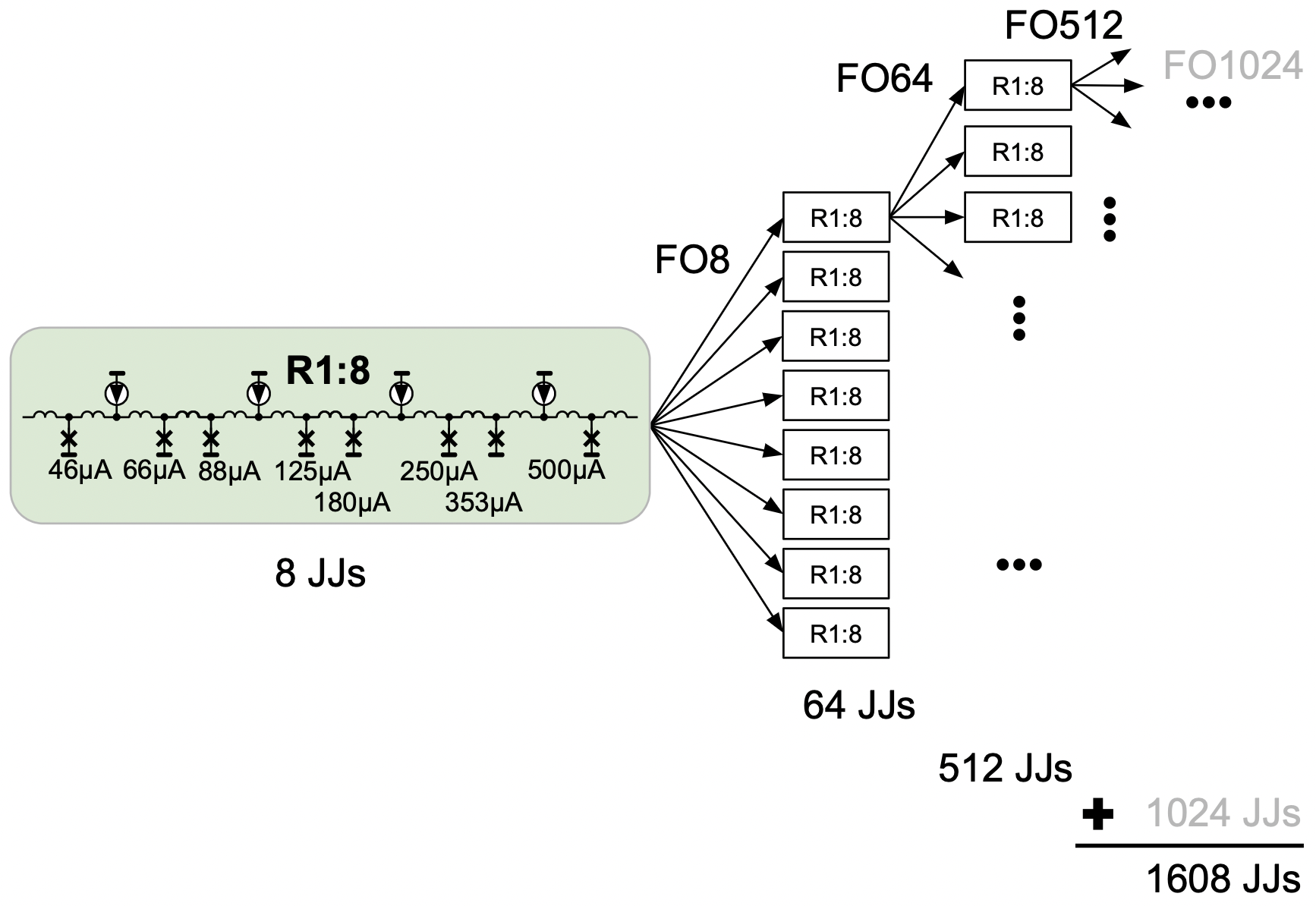}
    \caption{A FO1024 tree using ranking. Chains of JTLs followed by an FO8 serve as the modular building block and are depicted here as rectangles with the label R1:8. Each building block amplifies from rank-1 to rank-8, as shown in the first chain, and costs eight JJs. This balanced tree costs 1608 JJs, whereas a FO1024 tree using conventional splitters costs 3069 JJs---a savings of 47.6\%.}
    \label{fig:fo1000}
\end{figure*}

\begin{figure*}[h]
    \centering
    \includegraphics[scale=0.37]{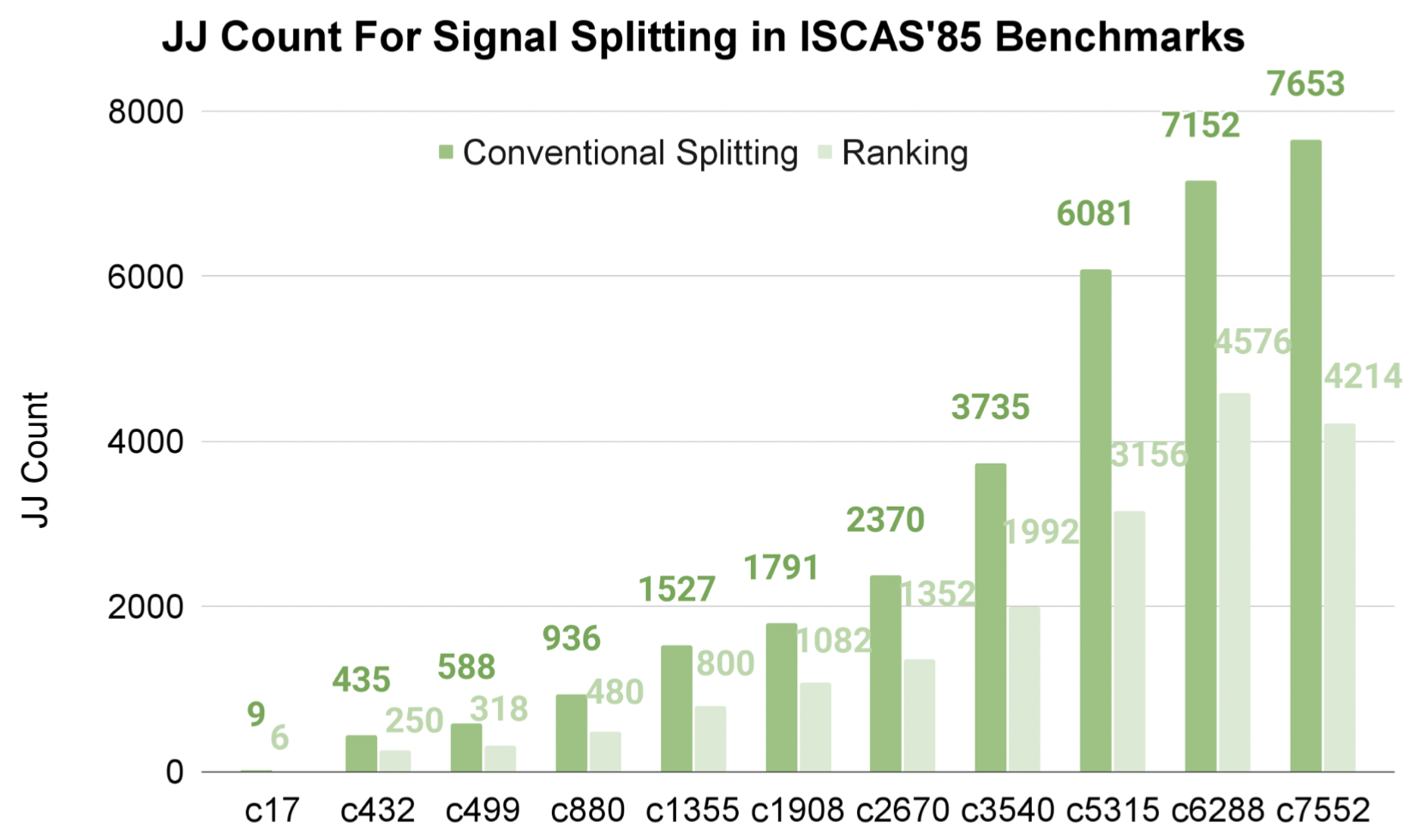}
    \caption{The \revised{number of JJs required for signal fan-out} in unmodified ISCAS'85 benchmark circuits is estimated for both conventional splitting and the proposed approaches. The average JJ savings with \revisedd{JJ sharing and cell ranking} is 43.3\%.
    }
    \label{fig:iscas85hist}
\end{figure*}

\begin{figure}
\begin{center}
\resizebox{0.7\columnwidth}{!}{%
\begin{tabular}{|c|c|c|c|c|c|c|}
\hline
\multicolumn{1}{|c|}{\textbf{Benchmark}} & \multicolumn{1}{c|}{\textbf{\begin{tabular}[c]{@{}c@{}}Improvement\\ (\revised{$p_a=\sqrt2$})\end{tabular}}} & \multicolumn{1}{c|}{\textbf{\begin{tabular}[c]{@{}c@{}}Improvement\\ (\revised{$p_a=2$})\end{tabular}}} \\ \hline
\textbf{c17}                                                                                           & 33.3\%                                                                                          & 33.3\%                                                                                               \\ \hline
\textbf{c432}                                                                                           & 42.5\%                                                                                       & 50.3\%                                                                                               \\ \hline
\textbf{c499}                                                                                           & 45.9\%                                                                                       & 65.3\%                                                                                               \\ \hline
\textbf{c880}                                                                                           & 48.7\%                                                                                       & 60.5\%                                                                                               \\ \hline
\textbf{c1355}                                                                                           & 47.6\%                                                                                       & 55.1\%                                                                                               \\ \hline
\textbf{c1908}                                                                                           & 39.6\%                                                                                       & 47.6\%                                                                                               \\ \hline
\textbf{c2670}                                                                                           & 43.0\%                                                                                      & 51.9\%                                                                                               \\ \hline
\textbf{c3540}                                                                                           & 46.7\%                                                                                      & 56.5\%                                                                                               \\ \hline
\textbf{c5315}                                                                                           & 48.1\%                                                                                      & 58.7\%                                                                                               \\ \hline
\textbf{c6288}                                                                                           & 36.0\%                                                                                      & 53.9\%                                                                                               \\ \hline
\textbf{c7552}                                                                                           & 44.9\%                                                                                      & 56.0\%                                                                                               \\ \hline
\textbf{Average}                                                                                           & 43.3\%                                                                                      & 53.6\%                                                                                               \\ \hline
\end{tabular}
}
\caption{Percent improvements to JJ count for data signal splitting using ranking with $\sqrt{2}$ and $2$ step sizes in ISCAS'85 benchmarks compared to the conventional splitter-based methodology.}\label{tab:benchmarkvalues}
\end{center}
\end{figure}

    \item \textbf{Step} \ding{204}. Keep target cells the same rank if they share a source cell. In this case, the penultimate cell in the first stage is forced to be rank-2, the second cell in the second stage is forced to be rank-2, and the two middle stages in the first stage are forced to be the same rank, which is still unknown. The fourth cell in the second stage is also assigned rank-2, based on the sharing that occurred in Step \ding{203}. The cells that have yet to be ranked at this point are the top three cells in the first stage and the third cell in the second stage.

    \item \textbf{Step} \ding{205}. To define the ranks of the remaining cells, we rely again on the table in \revised{Fig.}~\ref{fig:fanoutcost} while considering the ranks and fan-outs of the cells that surround the ones in question. Amplifying JTLs are inserted in this step to electrically reinforce the connections between cells. This fills in the remaining unranked cells in this example: the top three cells in the first stage are assigned rank-3. This rank then propagates to the third cell in the second stage. Finally, amplifying JTLs are inserted after the second and fourth cells in the second stage to meet the FO2 and FO3 requirements of the connection between rank-2 source cells and rank-2 target cells. It is also now clear that inputs A$_0$ and B$_0$ should be sourced from rank-4 cells, and A$_1$ and B$_1$ should be sourced from rank-3 cells.

    \item \textbf{Step} \textcolor[rgb]{0.76,0.48,0.63}{\ding{206}}. 
    The last step is to design the clock tree using chains of amplifying JTLs that meet the fan-out and target cell rank requirements for every synchronous stage. \revised{In the shown example, we start with a rank-8 cell that fans out to 8, and then amplify each line to rank-4. Six of these rank-4 lines fan out to three rank-2 cells each, which covers the rank-2 cells in every stage from the second to the fifth, while two are reserved for sharing amongst the remaining rank-2 and rank-3 cells in the first and second stages. To amplify from rank-2 to rank-3, one JTL is used per line, following the guidelines provided in Fig.~\ref{fig:fanoutcost}.}
\end{enumerate}

To verify the functional correctness of the resulting design, analog simulations in the \revisedd{Cadence design suite} are performed. The cells used are based on publicly available designs from Stonybrook~\cite{polonsky_rsfq_1995,likharev_semenov91,polonskyOR,mukhanov1991elements}. To satisfy the requirement for cells with flexible ranks, Stonybrook's designs are scaled to the required rank and individually tested before being stitched together. JTLs are inserted to enforce rank and prevent timing violations. An example waveform demonstrating the adder's functional behavior is shown in \revised{Fig.}~\ref{fig:ksasim}. The total JJ count for the design is \revised{317} and includes logic cells, DC-to-SFQ converters, and JTLs. \revised{The bias margins are \revisedd{simulated and found} to be $+19.2\%/-3.8\%$}.

To quantify the gains compared to a traditional approach, we reimplement the same 2-bit KSA design, but this time without the proposed \revisedd{JJ sharing} and \revisedd{cell} ranking techniques. The achieved results indicate that a total of \revised{385} JJs are needed for the traditional implementation, including \revised{114} JJs for splitting. In other words, the proposed methodology leads to \revised{17.7\%} JJ savings. \revised{The estimated bias margins of the traditional implementation are $+7.7\%/-23.1\%$, which are comparable to that of the rank-based version.}

\subsection{Modeling Larger Designs}

Our next task is to move beyond functional testing and quantify the benefits of the proposed approach on a larger scale. To this end, we first use a clock tree with FO1024 as an example case and then analyze ISCAS'85 benchmark circuits.

Regarding the FO1024 tree, rank-1 source and target cells are assumed. For the construction of the tree, R1:8 amplifying JTL chains are used in a way that extends \revised{Fig.}~\ref{fig:jtltest3} and expands upon the fan-out achieved in \revised{Fig.}~\ref{fig:jtltest4}. Our results indicate firstly that the design can be built with modularity, as the bias margins do not diminish beyond $+38.5\%/-46.2\%$. Secondly, the design, depicted in \revised{Fig.}~\ref{fig:fo1000}, uses 1608 JJs, while a similar tree with conventional splitters costs 3069 JJs, resulting in a savings of 47.6\%.

In the case of ISCAS'85 benchmarks~\cite{iscas85repo}, the fan-out requirements of each circuit are extracted by counting the fan-out for every signal in the corresponding Verilog file. The fan-out ranges from 1 to 16. In our analysis, we constrain fan-out per stage to FO8, as R1:8 JTL chains have shown to deliver satisfying bias margins. The JJ counts are shown in \revised{Fig.}~\ref{fig:iscas85hist} and percent improvements in \revised{Fig.}~\ref{tab:benchmarkvalues}. A rank-based approach grants an average JJ savings of 43.3\% for signal splitting, 32.3\% for clock fan-out, \revised{and 10\% for the total JJ count, even without taking into consideration the effects of path-balancing, which is expected to inflate these numbers further}.

\subsection{Increasing the Step Size}

Lastly, we analyze the effects of a parameter that so far has been constant: the step size. More specifically, a $\sqrt{2}$ amplification step size \revised{$p_a$} has been assumed in all prior cases, following the principles of early SFQ~\cite{likharev_semenov91} and microwave splitter design~\cite{wilkinson}. However, SPICE-level results indicate that a step of $2$ \revised{within each JTL \reviseddd{and a step of $\sqrt{2}$} between} can potentially work well in the case of R1:8 JTL chains\footnote{\reviseddd{The first JTL amplifies from rank-1 to rank-3 (corresponding to the ranks in Figure~\ref{fig:fanoutcost}), the second from rank-4 to rank-6, and the third from rank-7 to rank-8 to compensate for the asymmetry given by this step size and critical current range.}}. \reviseddd{This arrangement, simulated with the same test setup and termination as in} Section~\ref{eval}, \reviseddd{has bias margins of $+23.1\%/-23.1\%$, which are somewhat diminished from those of a rank-1 to rank-8 chain with a step of $\sqrt{2}$: $+42.3\%/-23.1\%$}. When combined with JJ \revisedd{sharing} and \revisedd{cell} ranking, the effects of \revised{intra-JTL $p_a$=2} on fan-out improvements to ISCAS'85 benchmarks are shown in \revised{Fig.}~\ref{tab:benchmarkvalues}. The results indicate an increase in average savings for signal splitting to 53.6\%.

\section{\revisedd{Design Trade-Off Discussion}}\label{tradeoffs}
\revisedd{The trade-off space between circuit area, delay, power, and reliability is intricate and cell ranking serves as a new control to steer optimizations. In Section~\ref{methods}B, three design methodologies incorporating cell ranking were presented. In this Section, we elaborate on the design trade-offs associated with each design methodology.}

\revisedd{In the first methodology, shown in Fig.~\ref{fig:designmeth1}, all logic cells feature the same baseline current, and therefore rank. Under this assumption, no change in existing cell designs and libraries is needed and splitting can be accomplished with cascaded amplifying JTLs. Note that JJ count instead of area is used to quantify resource efficiency improvements, as the provided results are simulation- and not fabrication-based. JJ count is frequently used as a first-order estimate of design complexity~\cite{massoud_morefanout,irds22} because it relates linearly to circuit area for closely-sized JJs and inductors, which is the case in this design methodology.}

\revisedd{In the second, exemplified by Fig.~\ref{fig:designmeth2}, the lowest rank is picked as the baseline. A low rank comes with higher switching speeds, lower power dissipation, and smaller JJs. Smaller JJs do not directly imply a smaller area per cell, as JJ size is inversely related to the size of shunt resistors and inductors~\cite{tolpygo2016superconductor}. However, JJ count remains an important indicator of overall resource efficiency because it tracks component count. Regarding circuit reliability---a valid concern when fabricating large-scale designs---cells with baseline currents below $\sim$50~$\mu$A may suffer higher bit error rates at 4.2 K~\cite{tolpygo2016superconductor}. That said, the applicability of JJ sharing and cell ranking is independent of the above-discussed current values; thus, the minimum acceptable baseline current value for a particular design and process technology is configurable at design time.}

\revisedd{Lastly, in the third case, depicted in Fig.~\ref{fig:designmeth3}, the goal is to demonstrate the potential benefits of incorporating full flexibility in the cell ranking arrangement. The drawback of this approach is that the size of the corresponding cell libraries increases as multiple copies of the same gate, each with a different baseline current, are needed---this is similar to the case of CMOS cell libraries that feature various driving configurations. The advantage is that area, speed, energy, reliability, and fan-out overhead are left as free variables for potential optimization. The critical current range and step size serve as additional degrees of freedom; for example, small or large baseline critical currents can be incorporated or ignored.}

\revisedd{In summary, each presented design methodology reveals a distinct set of relationships between rank and established metrics with various degrees of flexibility. The discussed trade-offs unwrap a plethora of optimization opportunities that are beyond the scope of this work. Nevertheless, the presented evaluation can function both as a high-level model and a good starting point for those optimizations while also providing strong evidence of the feasibility and foreseeable gains of JJ sharing and cell ranking.}

\section{Conclusion}

Advancements towards improving the computational density of superconductor systems commonly target the device, logic, and architecture levels. In this work, we take on a different approach and focus on \revised{cell abutment by} maximizing the utility of JJs already present in each logic cell. The key idea behind our efforts is that boundary JJs can also be used for splitting purposes, thereby eliminating frequently-used splitter cells that account for up to \revised{$\sim$30\%} of the total \revised{number} of JJs.

To facilitate this idea, we firstly proposed a JJ \revisedd{sharing} technique that examines the anatomy of existing SFQ cells and identifies JJs that are candidates for reuse. We secondly presented a ranking methodology that allows for the reliable stitching of SFQ cells with variable baseline critical currents in a way that abstracts analog design concerns. Our results indicate that such designs are functionally correct, reduce current consumption, and exhibit satisfying bias margins. \revised{Rank-based JJ \revisedd{sharing}} is also shown to deliver \revised{17.7\%} \revised{savings in the total JJ count} of a 2-bit KSA, an average 43.3\% savings for data signal splitting in ISCAS'85 benchmarks, and a 47.6\% savings for a fan-out tree with FO1024, compared to the same designs with conventional splitting techniques. Moreover, we notice that a change in the amplification step size, from $\sqrt{2}$ to $2$, can lead to an increase in average savings to 53.6\% for data signal splitting in ISCAS'85 benchmarks \reviseddd{at the cost of diminished bias margins}. The integrity of this setup was evaluated through transient simulations, and further investigation is needed to identify additional limitations on this parameter.

\revisedd{Lastly, the interplay between rank and various design metrics was analyzed in the context of three design methodologies. Each methodology offers different benefits and drawbacks, although greater flexibility in cell ranking tends to supply greater potential for optimization and, thus, greater potential for exploitation at the CAD-level.}

\section*{Acknowledgment}
This material is based upon work supported by the National Science Foundation under Grants No. 2006542 and 1763699 and the Under Secretary of Defense for Research and Engineering under Air Force Contract No. FA8702-15-D-0001. Any opinions, findings, conclusions or recommendations expressed in this material are those of the author(s) and do not necessarily reflect the views of the Under Secretary of Defense for Research and Engineering.


\bibliographystyle{IEEEtran}
\bibliography{refs}


\begin{IEEEbiographynophoto}{Jennifer Volk} is a Ph.D. student at the University of California, Santa Barbara researching how to best utilize the quirks of superconductor electronics to benefit circuit and architecture design. She received her M.S. degree in Electrical and Computer Engineering in 2021 from UC Santa Barbara and her B.S. degree in Electrical Engineering in 2016 from UC Santa Cruz.
\end{IEEEbiographynophoto}
\begin{IEEEbiographynophoto}{George Tzimpragos} is an Assistant Professor of Computer Science and Engineering at the University of Michigan. His research focuses on logic-architecture co-optimizations for emerging applications and devices. Prior to joining the University of Michigan, he received his Ph.D. in computer science from UC Santa Barbara (2022) and M.S. in electrical and computer engineering from UC Davis (2016). His alma mater is the National Technical University of Athens in Greece. 
\end{IEEEbiographynophoto}
\newpage
\begin{IEEEbiographynophoto}{Alex Wynn} received a B.A. degree in astronomy and physics from Boston University in 2012, and an M.A. degree in physics from Vanderbilt University in 2014. Since 2014, he has been working on the development of superconductor electronics technologies at MIT Lincoln Laboratory.
\end{IEEEbiographynophoto}
\begin{IEEEbiographynophoto}{Evan Golden} is a member of the technical staff at MIT Lincoln Laboratory and conducts research on superconducting electronics for classical and quantum computing. Before joining Lincoln Laboratory, he received a B.S. degree in engineering physics from University of Colorado Boulder in 2017 and performed research at the University of Colorado and the National Institute for Standards and Technology (NIST).
\end{IEEEbiographynophoto}
\begin{IEEEbiographynophoto}{Timothy Sherwood} is a Professor of Computer Science at UC Santa Barbara specializing in the development of computing systems exploiting novel technologies. He is a Fellow of the ACM and IEEE and currently serves as the Interim Dean of the College of Creative Studies. Prior to joining UCSB he received his B.S. at UC Davis (1998), and M.S. and Ph.D. from UC San Diego (2003) all in Computer Science and Engineering.
\end{IEEEbiographynophoto}

\end{document}